\documentclass[10pt,journal,compsoc]{IEEEtran}
\ifCLASSOPTIONcompsoc
  \usepackage[nocompress]{cite}
\else
  \usepackage{cite}
\fi
\ifCLASSINFOpdf
\else
\fi
\makeatletter
\newcommand{\maketitlenew}{\@maketitle}
\makeatother

\usepackage{algorithm}
\usepackage{algpseudocode}

\usepackage{amssymb}
\usepackage{amsmath}
\usepackage{amsfonts}       
\usepackage{mathtools}
\usepackage{nicefrac}       
\usepackage{amsthm}         
\usepackage{bbm}
\usepackage{bm}
\allowdisplaybreaks
\newtheorem*{thm*}{Theorem}
\newtheorem*{defn*}{Definition}
\newtheorem*{lem*}{Lemma}
\newtheorem*{prf*}{Proof}
\newtheorem*{cor*}{Corollary}
\newtheorem*{prop*}{Proposition}
\newtheorem*{exmp*}{Example}

\usepackage{graphicx}
\graphicspath{ {figures/} }
\usepackage{epsfig}
\usepackage{subcaption}
\usepackage[export]{adjustbox}

\usepackage[utf8]{inputenc} 
\usepackage[T1]{fontenc}    
\usepackage{microtype}      
\usepackage{textcomp}

\usepackage{booktabs}       
\usepackage{makecell}

\usepackage{color}
\usepackage{xcolor}
\usepackage{xkeyval}

\usepackage{tikz}
\usetikzlibrary{bayesnet}
\usetikzlibrary{positioning}
\usetikzlibrary{arrows}
\usetikzlibrary{calc}

\usetikzlibrary{positioning,fit,shapes,backgrounds,circuits.logic.US}

\tikzstyle{server}=[circle, line width=0.5pt, rounded corners=0.1mm, draw=black!100, fill=tud3a!100]
\tikzstyle{vertex}=[circle, line width=1.5pt, draw=tud0d, fill=white]
\tikzstyle{dispatcher} =[and gate US, line width=0.5pt, draw=black!100, fill=tud0c!100]
\tikzstyle{dotbox} = [draw=white, fill=white, rectangle,  inner sep=10pt, inner ysep=20pt]

\tikzset{three_sided/.style={
		draw=none,rectangle, 
		append after command={
			[shorten <= -0.5\pgflinewidth]
			([shift={(-1.5\pgflinewidth,-0.5\pgflinewidth)}]\tikzlastnode.north west)
			edge([shift={( 0.5\pgflinewidth,-0.5\pgflinewidth)}]\tikzlastnode.north east)
			([shift={( 0.5\pgflinewidth,-0.5\pgflinewidth)}]\tikzlastnode.north east)
			edge([shift={( 0.5\pgflinewidth,+0.5\pgflinewidth)}]\tikzlastnode.south east)
			([shift={( 0.5\pgflinewidth,+0.5\pgflinewidth)}]\tikzlastnode.south east)
			edge([shift={(-1.0\pgflinewidth,+0.5\pgflinewidth)}]\tikzlastnode.south west)
		}
	}
}

\usepackage{paralist} 
\usepackage[shortlabels]{enumitem}
\usepackage{amstext} 
\usepackage{array} 
\usepackage{url}            
\usepackage{xspace}
\usepackage{csquotes}
\usepackage{etoolbox}
\usepackage{comment}
\usepackage{hyperref}

\usepackage{url}
\usepackage[capitalise]{cleveref}
\usepackage{cite}
\def\BibTeX{{\rm B\kern-.05em{\sc i\kern-.025em b}\kern-.08em
    T\kern-.1667em\lower.7ex\hbox{E}\kern-.125emX}}

\usepackage{ifthen}
\usepackage{balance}

\usepackage{todonotes}
\newboolean{todo}
\setboolean{todo}{true} 

\newboolean{longVersion}
\setboolean{longVersion}{false}  
\newcommand{\IfLongVersion}[1]{\ifthenelse{\boolean{longVersion}}{#1}{}}
\newcommand{\IfShortVersion}[1]{\ifthenelse{\boolean{longVersion}}{}{#1}}

\usepackage{blindtext}
\newboolean{blind}
\setboolean{blind}{false} 
\newcommand{\blindparagraph}[2][]{\ifthenelse{\boolean{blind}}{\blindtext[1]}{}}

\usepackage{hyphenat} 

\DeclareMathOperator*{\argmax}{\arg\max}

\DeclareMathOperator{\E}{\mathsf{E}\,}

\DeclareMathOperator{\GamDis}{\mathrm{Gam}\,}
\DeclareMathOperator{\GeomDis}{\mathrm{Geom}\,}
\DeclareMathOperator{\ExpDis}{\mathrm{Exp}\,}

\DeclareMathOperator{\BinDis}{\mathrm{Bin}\,}

\DeclareMathOperator{\PoisDis}{\mathrm{Pois}\,}

\DeclareMathOperator{\TPDis}{\mathrm{TP}\,}
\DeclareMathOperator{\BetaDis}{\mathrm{Beta}\,}

\usepackage[nolist,nohyperlinks]{acronym}
\begin{acronym}[ABC]
\acro{pos}[POS]{Partial Observability Scheduler}
\acro{posmdp}[POSMDP]{Partially observable semi Markov decision process}
\acro{pomdp}[POMDP]{partially observable  Markov  decision  process}
\acro{mdp}[MDP]{Markov  decision  process}
\acro{mcts}[MCTS]{Monte Carlo tree search}
\end{acronym}

\newcommand{\mcts}{\ac{mcts}\xspace}
\newcommand{\pomdp}{\ac{pomdp}\xspace}
\newcommand{\mdp}{\ac{mdp}\xspace}
\newcommand{\posmdp}{\ac{posmdp}\xspace}

\begin{document}
%
\title{Load Balancing in Compute Clusters with Delayed Feedback}

%
%
%
%

\author{
Anam~Tahir,
Bastian~Alt,
Amr Rizk,~\IEEEmembership{Senior Member,~IEEE,}
Heinz~Koeppl
\IEEEcompsocitemizethanks{\IEEEcompsocthanksitem A. Tahir, B. Alt and H. Koeppl are with the Self-Organizing Systems Lab (SOS), Department of Electrical Engineering and Information Technology, Technische Universität Darmstadt, Darmstadt, Germany. E-mail: {anam.tahir, bastian.alt, heinz.koeppl}@tu-darmstadt.de.\protect\\
\IEEEcompsocthanksitem A. Rizk is with the Communication Networks and Systems Lab, Universität Duisburg Essen, Germany. E-mail: amr.rizk@uni-due.de.}
}

\IEEEtitleabstractindextext{
\begin{abstract}

Load balancing arises as a fundamental problem, underlying the dimensioning and operation of many computing and communication systems, such as job routing in data center clusters, multipath communication, Big Data and queueing systems. In essence, the decision-making agent maps each arriving job to one of the possibly heterogeneous servers while aiming at an optimization goal such as load balancing, low average delay or low loss rate. One main difficulty in finding optimal load balancing policies here is that the agent only partially observes the impact of its decisions, e.g., through the delayed acknowledgements of the served jobs. 
In this paper, we provide a partially observable (PO) model that captures the load balancing decisions in parallel buffered systems under limited information of delayed acknowledgements. 
We present a simulation model for this PO system to find a load balancing policy in real-time using a scalable Monte Carlo tree search algorithm. 
We numerically show that the resulting policy outperforms other limited information load balancing strategies such as variants of Join-the-Most-Observations and has comparable performance to full information strategies like: Join-the-Shortest-Queue, Join-the-Shortest-Queue(d) and Shortest-Expected-Delay. Finally, we show that our approach can optimise the real-time parallel processing by using network data provided by Kaggle.

\end{abstract}

}

\maketitle
\IEEEdisplaynontitleabstractindextext

%
\IEEEpeerreviewmaketitle

\IEEEraisesectionheading{\section{Introduction}\label{sec:introduction}}

\IEEEPARstart{A}{s} the growth rate of single-machine computation speeds started to stagnate in recent years, parallelism seemed like an effective technique to aggregate the computation speeds of multiple machines.
Since then, parallelism has become a main ingredient in compute cluster architectures~\cite{Dean:2008:MSD,2014:CVH} as it incurs less processing and storage cost on any one individual server \cite{griffith2009above}.

Beyond the aggregation of capacity, a major difficulty in the operation of parallel servers is optimizing for low latency and loss.
The key to this optimization is the mapping of the input, that we denote as \emph{jobs}, to the different and possibly heterogeneous serving machines, denoted \emph{servers}, of time-varying capacity and finite memory (buffers). This mapping of input to servers is carried out by a decision-making agent we refer to as the \emph{load-balancer}.

Classical results prove the optimality of the Join-the-Shortest-Queue (JSQ) algorithm in terms of the expected job delay\cite{hordijk1990optimality,winston1977optimality} when the servers are homogeneous, have infinite buffer space and the job service times are independent, identically (i.i.d) and exponentially distributed. For the case when the service times are exponentially distributed but with different service rates, Shortest-expected-delay (SED)  has been shown to minimize the mean response time of jobs, especially in the case of heavy traffic limits \cite{banawan1989load,selen2016steady}.

Note, however, that such types of algorithms assume that the decision-making agent has accurate, timely and synchronized information of all servers and their queues.
\emph{In practice this assumption does not hold}, e.g., in data center clusters it cannot be assumed that a load-balancer has timely and synchronized information of all available servers but rather observes some event or time-triggered server feedback.

The goal of this work is to model and optimize the load balancing decision-making over parallel and heterogeneous servers having finite buffers and each following first-in-first-out (FIFO) order, where these servers provide randomly delayed acknowledgements back to the load-balancer.
These acknowledgments are sent by the servers to inform the load-balancers of the \textit{number of jobs} the servers have processed in the last epoch.
This model captures that the load-balancer is only able to partially observe the server states at the decision time points. 
The load-balancer does not directly observe the server queues, rather it receives acknowledgements for completed jobs that are randomly delayed on the way back to it.
Our contributions are:
\begin{itemize}
\item We present a model for controlled load balancing in parallel buffered systems with randomly delayed acknowledgements.
\item  We find a control law by optimizing a predefined objective function subject to the stochastic dynamics induced by the model.
\item We present \emph{POL} - a Partial Observability Load-Balancer, which maps incoming jobs to the parallel servers. POL estimates the unknown parameters of the parallel system at runtime and despite partial observations achieves a job drop rate and response time comparable with full information load-balancers. The performance evaluation is carried out in a simulation environment. Note that POL can be used for any number of servers and for any kind of inter-arrival and service time distributions, and we have tested a few of them.
\item Lastly, we also show how POL can optimize parallel processing in clusters by using real network data provided by Kaggle \cite{kaggle, kaggle2}.
\end{itemize}

The rest of this paper is organized as follows: In \cref{sec:related_work} we first discuss the related work. Then,  in \cref{sec:system-model} we outline the system model and give some background on the key topics of this work. In \cref{sec:queueing_system_pomdp}, we give our contributions starting from the modelling of the partially observable queueing system to our Monte Carlo approach for finding a near-optimal load balancing policy.
In \cref{sec:sim_results},  we give our simulation results and discuss the inference of unobserved system parameters in \cref{sec:inference} along with an experiment with real world data. And then we conclude the paper in \cref{sec:conclusion}. 

\section{Related Work}
\label{sec:related_work}

\looseness=-1 Dynamic load balancing for the performance optimization of parallel systems has fueled numerous seminal algorithms such as Join-the-shortest-queue (JSQ), 
and Shortest-expected-delay (SED), and more generally Power-of-d policies \cite{banawan1992comparative, van2018scalable, whitt1986deciding}.
JSQ provides optimal decisions, for minimizing the mean response time of a job, when the servers are homogeneous and the service times are independent and identically exponentially distributed \cite{hordijk1990optimality,winston1977optimality}.
A similar approach for heterogeneous servers is the SED algorithm, which implicitly considers the server rates and maps an incoming job to the server which provides the smallest expected response time for the job at hand.
SED is known to perform well for heterogeneous servers, especially in the case of heavy traffic \cite{banawan1989load, selen2016steady}.
 
However, when the number of parallel systems $N$ becomes large, the assumption of knowing the state of the entire system before every decision becomes too strong.
The state may be the queue length or the required cumulative service times for the waiting jobs at each system.
Depending on the type of system this information, e.g., the service times for servers of random varying capacity, is not known in advance.
Control theory and the recently attractive machine learning approaches have also been extensively applied to stochastic queuing networks in order to analyse their performance as self-adaptive software systems, see \cite{arcelli2020exploiting,shevtsov2017control} and references therein.

Power-of-d policies provide a remedy to the problem of the decision-making agent not being able to know all system states at decision time. Here, a number of servers, specifically $d<N$, is repeatedly polled at random at every decision instant and hence, JSQ(d) or SED(d) is performed on this changing subset~\cite{963420}.
This policy is enhanced by a short term memory that keeps knowledge of the least filled servers from the last decision instant~\cite{shah2002use, psounis2002efficient}.
Hence, instead of choosing $d$ servers at random for every job, the decision is based on a combination of new randomly chosen $d$ servers and the least filled servers known from the last decision.
For our evaluation purposes we have chosen to compare our algorithm with SED, JSQ and JSQ(d) since the comparison is with respect to the classes of full and limited information at the load-balancers. 

\emph{The strong assumption behind the different variants of JSQ, SED and the Power-of-d policies is that at every decision instant the load-balancer is aware of the current system state of all or some of the servers}. 
The main difference in this work is that we hypothesize that this assumption of instant (and full) knowledge is often not realistic, as due to the distributed nature of the system the load-balancer may only observe the impact of a decision that maps a job to a server after some non-deterministic feedback time.
This feedback time may arise, e.g., due to the propagation delay or simply because the decision-making agent receives feedback only after the job has been processed.
The impact of this non-deterministic and heterogeneous feedback time on the decision-making process is significant, as the consequence of a  decision on performance metrics such as job response times or job drop rates is only partially observed at the decision instant.

Markov decision processes, MDP, have been used to model queuing systems and achieve optimal control under static and dynamic environments \cite{liu2019reinforcement, ayesta2022reinforcement}. In an MDP the current feedback, delayed or not, is known to the agent \cite{Sutton1998}. 
The concept of transitions between mechanisms to achieve better overall performance in a dynamic communication system, modelled as an MDP, has recently been proposed in \cite{alt2019transitions}.
The authors of \cite{winston1977optimality} used an MDP formulation together with a stochastic ordering argument to show that JSQ maximizes the discounted number of jobs to complete their service for a homogeneous server setting.
The authors of \cite{krishnan1987joining} study the problem of allocating customers to parallel queues.
They model this problem as an MDP with the goal of minimizing the sojourn time for each customer and produce a 'separable rule', which is a generalization of JSQ, for queues with heterogeneous servers (in rates and numbers).
Note, however, that it is assumed that the queue filling is known and available without delay to the decision-making agent.

In \cite{altman1992closed} the authors assume that the decision-making agent receives the exact queue length information but with a delay of $k$ steps. 
They formulate their system as a Markov control model with \textit{perfect state information} by augmenting to the state space the last known state (exact queue length) and all the actions taken until the next known state.
In their work, they solve the flow control problem by \textit{controlling the arrival} to a single server queue and show for $k=1$ that the optimal policy is of threshold type and depends on the last action. 
In \cite{kuri1995optimal}, along with the single server flow control problem with similar results as \cite{altman1992closed}, they also showed that when $k=1$, the optimal policy for minimizing the discounted number of jobs in a system of two parallel queues is join-the-shortest-expected-length.
In \cite{artiges1993optimal}, the decision-making agent at time \textit{n}, \emph{knows the number of jobs} that were present in each (infinite) queue at time \textit{n-1}, such that it takes decisions at a deterministic delay of one time slot.
The state space at any time \textit{n} is augmented and contains the actual queue filling at time \textit{n-1}, the action taken at time \textit{n} and if there was an arrival at time \textit{n}.

In our work, we assume that not only is the information received by the load-balancer randomly delayed, but it is also not the state of the buffers, rather it is in the form of acknowledgements of the number of jobs processed.
This makes the system \textit{partially observable} and complex to solve, i.e., in the sense of a \pomdp model.
Several online and offline algorithms have been introduced to solve a \pomdp \cite{chades2021primer}, but there is little work on optimizing queuing systems as a \pomdp.
Standard solutions, for a \pomdp, that do full-width planning \cite{ross2008online}, like Value iteration and Policy iteration, perform poorly when the state space grows too large. This can easily be the case in queuing systems, due to the curse of dimensionality and the curse of history \cite{kaelbling1998planning}, \cite{pineau2011anytime}. For such large state problems, the POMCP algorithm \cite{silver2010monte}, which uses the  \mcts approach, is a fast and scalable algorithm for solving a \pomdp. 
Our algorithm also makes use of the \mcts algorithm specifically designed to simulate a parallel queueing system with finite buffers. In addition, we also designed a Sequential Importance Resampling (SIR) particle filter to deal with the delayed feedback acknowledgements in the queueing system. 
We also use this MCTS approach for solving a \posmdp \cite{yu2006approximate}, which is needed when the time between decision epochs is no more exponential.

\section{System Model}
\label{sec:system-model}
We consider a queuing system with $N$ parallel servers each having its own finite FIFO queue.
The queue filling is denoted $b_i \in \mathcal{B}_i, i=1,\dots,N$,
where $\mathcal{B}_i=\{0,\dots,\bar{b}_i\}$ and
$\bar{b}_i$ is the buffer space for the $i$-th queue. We define the vector of queue sizes as $\mathbf b =[b_1,\dots,b_N]^\top$. In the following, we will use boldface letters to denote column vectors.

\begin{figure}
\center
\includegraphics[scale=.16]{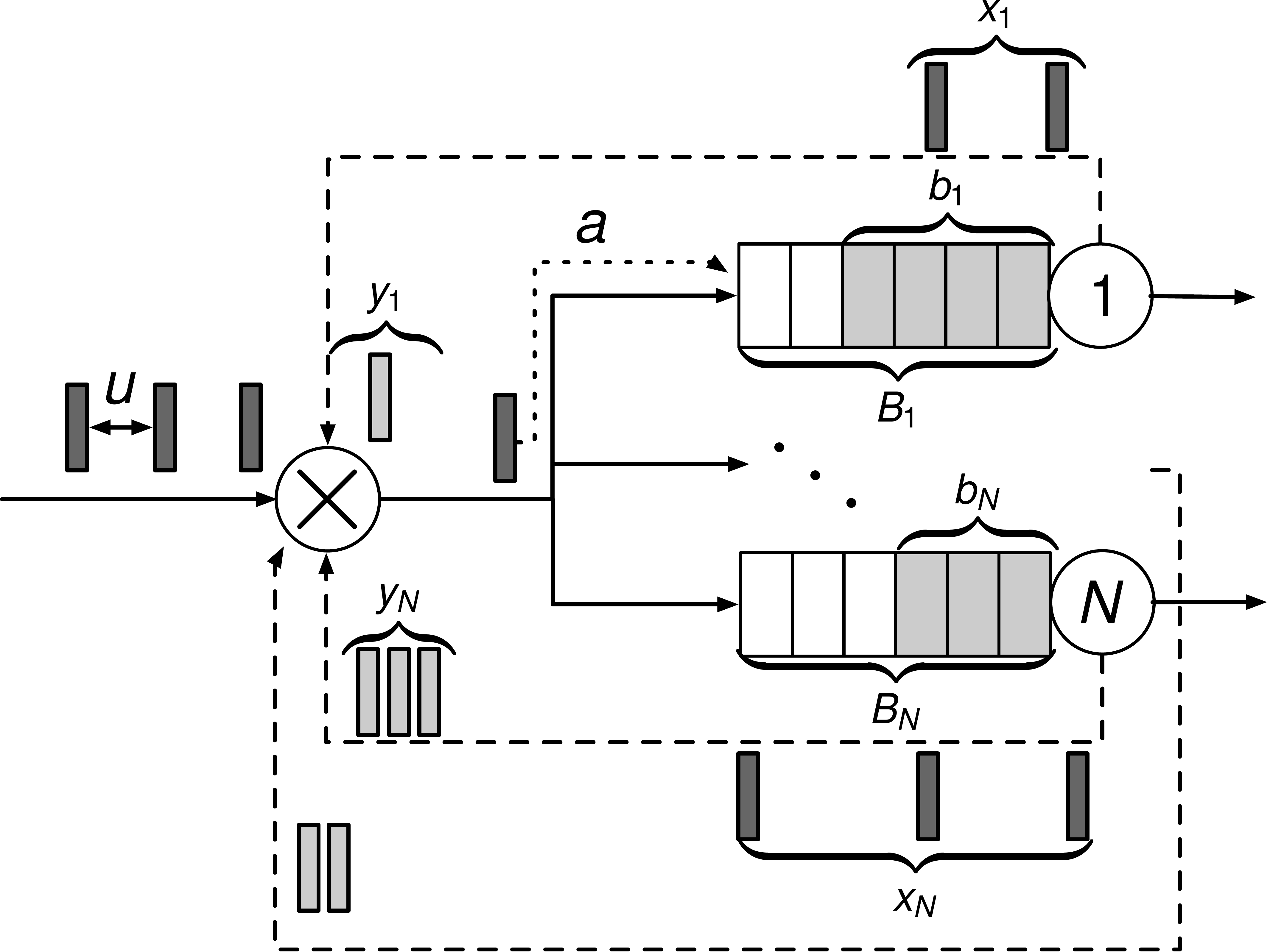}
\caption{A parallel queuing system with a load-balancer that maps jobs to servers.
The load-balancer observes the inter-arrival times $u$ and the feedback, i.e., the number of acknowledgements, from each server $y_i$. The load-balancer \emph{does not observe} the queue states $b_i$, the job service times, or the delayed feedback $x_i$, i.e., the number of acknowledgements on the way back.}
\label{fig:queuein_dynamics}
\vspace{-15pt}
\end{figure}

We consider the general case of heterogeneous servers
where the service times, $V_i^{(1)},V_i^{(2)}\dots,$ of consecutive jobs at the $i$-th server\footnote{We denote random variables by uppercase letters and their realizations as lowercase letters.} are independent and identically distributed (i.i.d) according to a probability density $f_{V_i}$.
Homogeneous jobs arrive to the load-balancer, as depicted in \cref{fig:queuein_dynamics},
according to some renewal process described by the sequence $(T^{(n)})_{n\in \mathbb{N}}$, where the job inter-arrival times $ U^{(j)}:=T^{(j+1)} - T^{(j)}$ are drawn i.i.d from a distribution $F$ leading to an average arrival rate $\lambda$.
Each arriving job is mapped by the load-balancer to exactly one server, where the service rate for that job is random and we denote the average service rate of the $i$-th server as~$\mu_i$.
A job that is mapped to a full buffer is lost.
When a job leaves the system, the server sends an acknowledgement back to the load-balancer, to inform it of a slot getting free in its associated queue.
The load-balancer then uses this feedback acknowledgement from each server to  calculate the current buffer fillings, $b_i$, of each queue.

A major challenge for deciding on the job routing arises when the acknowledgments from the server are delayed.
Here, we incorporate three main delay components, i.e., $(i)$ the job waiting time in the queue it was assigned to, $(ii)$ the job processing time, and $(iii)$ the propagation delay of the acknowledgement back to the load-balancer.
The third component \textit{makes the decision problem particularly hard} as a decision does not only impact the current state of the system but also future states due to its delayed acknowledgement feedback.
Note that the load-balancer  makes  the decision  based  only  on  these observed acknowledgments, thus making the system state partially observable (PO). 
We denote by $y_i$ the number of acknowledgements from the $i$-th server that are observed by the load-balancer in one inter-arrival time.
And we denote by $x_i$ the delayed feedback, which is the number of acknowledgments that are on the way back to the load-balancer but have not reached it yet.
Since the load-balancer is PO, it does not directly observe the (1) queue states $b_i$, (2) the job service times $v_i$, or (3) the delayed feedback $x_i$.
See Fig.~\ref{fig:queuein_dynamics} for further visualization.

Depending on the distribution type of the inter-arrival times, $U$, and service times, $V$, we model the decision-making process in this PO queuing system as a \emph{partially observable Markov decision process}, \pomdp, or a \emph{partially observable semi Markov decision process}, POSMDP.
A \pomdp has an underlying \mdp, where the actual state of the system is not known to the agent, i.e., the load-balancer in our case. This modelling can be used when both $U$ and $V$ are exponentially distributed, keeping the system Markovian.
For non-exponential $U$ and/or $V$, \posmdp formulation can be used, where the underlying process is now semi-Markov and the actual state of the system is still not known to the load-balancer. 

\subsection*{Markov Decision Process with Partial Observability}
\label{subsec:primer_pomdp}
An \mdp ~\cite{Sutton1998} is defined as a tuple $\langle\mathcal{S},\mathcal{A}, \mathcal{T}, \mathcal R \rangle$, where~$\mathcal{S}$ is the countable state space,
$\mathcal{A}$ is the countable action space,
$\mathcal{T}: \mathcal{S} \times \mathcal{S} \times \mathcal{A} \rightarrow [0,1]$ is the transition function,
$\mathcal R: \mathcal{S} \times \mathcal{S} \times \mathcal{A} \rightarrow \mathbb R$ is the reward function.
Throughout this work we assume time homogeneity for the transition function~$\mathcal T$, observation function~$\mathcal O$ and reward function~$\mathcal R$.
An MDP with partial observations is also a controlled Markov process, where the exact state of the process is latent.
It is in addition defined using; $\mathcal{Z}$, the countable observation space, and $\mathcal{O}: \mathcal{Z} \times \mathcal{S} \times \mathcal{A} \rightarrow [0,1]$, the observation function.
Consider the case of discrete epochs at time points $t\in \mathbb{N}_0$.
Note that the clock given by $t$ is an event clock and not a wall-clock time. 
If the decision epochs $t$ are exponentially distributed, then this PO process can be modelled as a \pomdp \cite{kaelbling1998planning}.
However, if $t$ is not exponentially distributed then under the condition that the decision-making is done only at the epochs $t$, it can be modelled as a \posmdp \cite{yu2006approximate}.

We consider a latent process $S^{(t)} \in \mathcal{S}$ that can be controlled by actions $A^{(t)} \in \mathcal{A}$.
Since, the state is latent, only observations $Z^{(t)} \in \mathcal{Z}$ are available.
The transition function $\mathcal{T}(s^\prime, s, a) := \mathbb P (S^{(t+1)}=s^\prime \mid S^{(t)}= s, A^{(t)}= a)$ is the conditional probability of moving from state $s$ under action $a$ to a new state ${s}^\prime$.
The observation function $\mathcal{O}(z,  s, a):=\mathbb P (Z^{(t+1)}=z \mid  S^{(t+1)} = s', A^{(t)}=a)$ denotes the conditional probability of observing ${z}$ under the latent state $ s'$ and action $a$. An agent receives a reward $R^{(t)} = \mathcal{R}(S^{(t+1)},S^{(t)},A^{(t)})$, which it tries to maximize over time.

Since the current state is not directly accessible by the agent, it has to rely on the action-observation history sequence, $\mathcal H^{(t)} = \{ A^{(0)},  Z^{(0)}, A^{(1)},  Z^{(1)}, \dots, A^{(t)},  Z^{(t)}\}$, up to the current time point $t$. A policy $\pi_t(a,h):=\mathbb P (A^{(t)}=a \mid \mathcal H^{(t)}=h )$ is the conditional probability of choosing action $a$ under action-observation history $h$.
The solution then corresponds to a policy which maximizes an objective over a prediction horizon.
The policy is defined as a function of the observation-action history of the agent, which makes it very challenging, since a naive planning algorithm requires to evaluate an exponentially increasing number of histories in the length of the considered time horizon. For this reason, different solution techniques are required.

Since keeping a record of the entire history, $h$, is not feasible, one way is to represent this history in terms of the belief state, $\bm{\rho}^{(t)} \in \Delta^{ \mathcal |S|}$, where $\Delta^{ \mathcal |S|}$ is an $\mathcal S$ dimensional probability simplex, $\boldsymbol \rho^{(t)}=[\rho_1^{(t)},\dots,\rho_{\mathcal |S|}^{(t)}]^\top$ and the components $\rho_s^{(t)}$ are the filtering distribution $\rho_s^{(t)}=\mathbb{P}(S^{(t)}=s \mid \mathcal{H}^{t}=h)$.
If the state space is huge, this will be a very high dimensional vector.
So, in order to break the curses of \textit{history} and \textit{dimensionality}, a certain number of particles can be used to represent the belief state $\bm{\rho}^{(t)}$ of the system at time $t$,
see \cite{silver2010monte} for further details.
These particles represent the belief state $\bm{\rho}^{(t)}$ of the system and are updated using Monte Carlo simulations based on the action taken and observations received. 
This is the approach that we build upon in this paper.
We consider an infinite horizon objective, where the optimal policy
 $\pi^*$ is found by maximizing the expected total discounted future reward
$\pi^* =  \argmax_{\pi}\sum^{\infty}_{t=0}  \E_{\pi}[\gamma^t R^{(t)}]$,     with $\gamma<1$.
Note that our work can easily be reformulated into finite horizon objectives.

\section{Load Balancing in Parallel Queuing Systems with Delayed Acknowledgments}
\label{sec:queueing_system_pomdp}
In this section, we explain how we model our partially observable (PO) system and then propose our solution.
\subsection{Modelling Load Balancing in the Partially Observable Queuing System }
\label{subsection:queue_as_pomdp}
In order to model the load balancing decision in a system with~$N$ parallel finite buffer servers (cf. Fig.~\ref{fig:queuein_dynamics}), as a PO process, we define the system state $\mathbf s\in \mathcal S$, using $\mathbf s=[\mathbf s_1,\dots, \mathbf s_N]^\top$. 
Here, $\mathbf s_i$ is the augmented state of the $i$-th queue that is defined as $\mathbf s_i= [b_i, x_i,y_i]^\top$, where,
$b_i \in \mathcal{B}_i$ denotes the current buffer filling at queue $i$, 
$x_i \in \mathcal{B}_i$ denotes the number of delayed acknowledgements for the jobs executed by the server $i$ but not observed by the load-balancer in the current epoch\footnote{An epoch corresponds here to one inter-arrival time.}, and 
$y_i \in \mathcal{B}_i$ denotes the number of acknowledgements actually \textit{observed} by the load-balancer in the current epoch.
Hence, the state space is $\mathcal S \subseteq {\mathbb{N}_0}^{3 N}$ and
an action $a \in \mathcal A$, with $\vert \mathcal A \vert=N$ corresponds to sending a job to the $a$-th server.
An observation $\mathbf z \in \mathcal Z$ is the vector of observed acknowledgements at the load-balancer, with the observation space being $\mathcal Z \subseteq {\mathbb N_0}^N$.

\subsection{The Dynamical Model}
\begin{figure}
\center
\includegraphics[scale=.55]{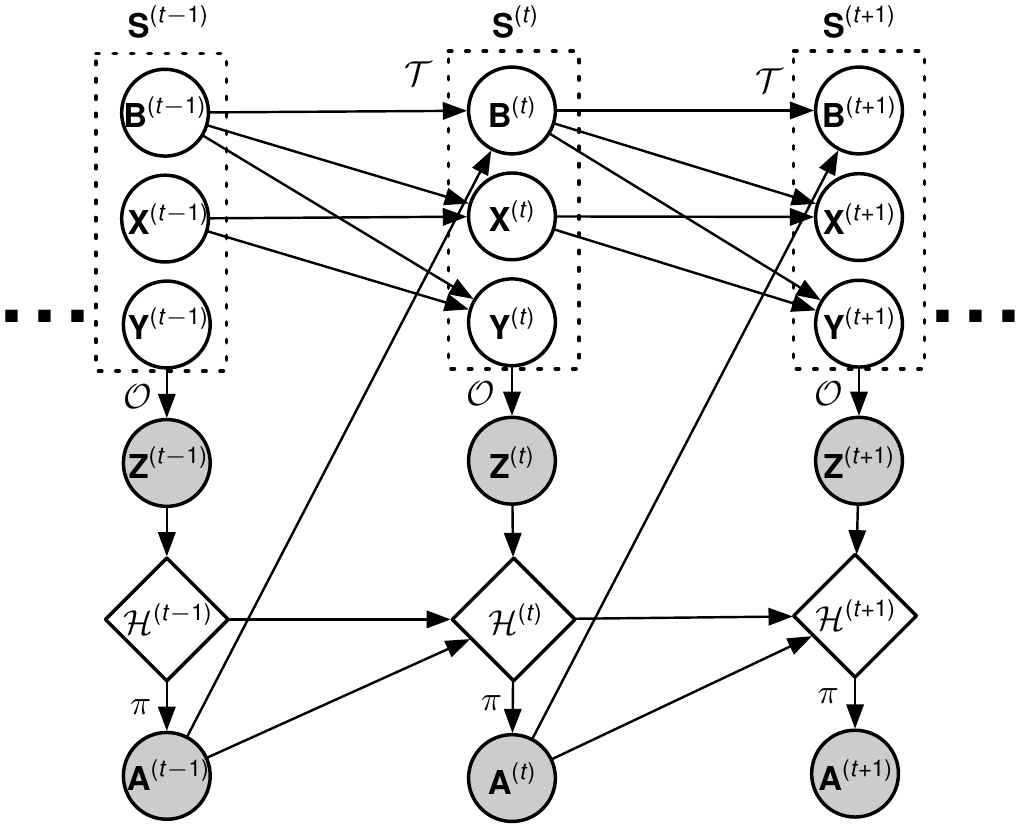}
\caption{Probabilistic graphical model of the partially observable  queuing system with delayed acknowledgements. Shown are three time slices, where grey nodes depict observed quantities and diamond shaped nodes denote deterministic functions.}
\vspace{-10pt}
\label{fig:pgm}
\end{figure}

Next, we describe the dynamics of the underlying processes of the PO model. 
The corresponding probabilistic graphical model is depicted in \cref{fig:pgm}.
In case of the \pomdp model, the time $t$ in \cref{fig:pgm} is exponentially distributed, while for \posmdp it can be random (non-exponential). 
As we are using Monte Carlo simulations to solve the PO system, the transition probabilities do not have to be defined explicitly. Therefore, we define the transition function indirectly as a generative process.

We consider a load-balancer (our decision-making agent) that makes a mapping decision at each job arrival, where the inter-arrival times $U^{(j)}$ are i.i.d.
In order to characterize the stochastic dynamics, we determine the random behaviour of the number of jobs $\tilde{k}_i$ that leave the $i$-th queue during an inter-arrival time. Note that the number of jobs leaving the queue is constrained by the current filling of the queue, $b_i$, of the $i$-th server, hence we use $\tilde{k}_i=\min (k_i, b_i)$, where $k_i$ is the number of jobs that \emph{can be served}, which is determined by the inter-arrival time and the service times of the $i$-th server.
with maximum capacity, $\bar b_i$.
Therefore, we define the generative model for the queuing dynamics as:
\begin{equation}
\mathbf{b}^\prime= \min (  \max (\mathbf{b}-\mathbf{k},\mathbf 0) +\mathbf{e}_{a}, \bar{\mathbf{b}}),
\end{equation}
where $\mathbf{b}$ denotes the queue size vector of the queuing system at some arrival time point and  $\mathbf{b}^\prime$ is the queue size vector of the queuing system at the next epoch.
By $\mathbf{k}$ we denote the non-truncated vector of number of jobs that can be served and $\mathbf{e}_a$ is a vector of all zeros, except the $a$-th position is set to one to indicate a mapping of the incoming job to the $a$-th server.
We use $\bar{\mathbf{b}}$ as  the vector of maximum buffer sizes, and $ \min (\cdot, \cdot)$ and $ \max (\cdot, \cdot)$ denote the element-wise minimum and maximum operation.

As the \emph{load-balancer only observes the job acknowledgements}, we update the augmented state space, $\mathbf{s}$ (as defined in Section \ref{subsection:queue_as_pomdp}), using the following stochastic update equation:
\begin{equation}
 \mathbf s^\prime
=
\begin{bmatrix}
\mathbf b^\prime\\
\mathbf x^\prime\\
\mathbf y^\prime
\end{bmatrix}
=
\begin{bmatrix}
 \min (  \max (\mathbf{b}-\mathbf{k},\mathbf 0) +\mathbf{e}_{a}, \bar{\mathbf{b}})\\
 \min( \mathbf b, \mathbf k) + \mathbf x - \mathbf l\\
\mathbf l
\end{bmatrix},
\label{eq:augmented_state}
\end{equation}
where $\mathbf{l}$ is the vector containing the number of jobs which are observed by the load-balancer for each queue at the current epoch.
The number of unacknowledged jobs from the previous epoch is denoted $\mathbf{x}$, which is updated by removing the observed jobs $\mathbf l$ and adding the newly generated acknowledgments given by $\min( \mathbf b, \mathbf k)$.
The delay model according to which $\mathbf{l}$ is calculated is given next.
\subsection{Delay Model}
\label{subsection:delay_model}
We assume that the number of jobs that can be served in one inter-arrival time is distributed as $K_i \sim f_{K_i}(k_i)$ and we choose a delay model, where
\begin{equation}
L_i  \mid b_i, k_i, x_i \sim \BinDis(\min( b_i,  k_i) +  x_i, p_i ).
\label{eq:binomial}
\end{equation}
Here, $\min( b_i,  k_i)$ is the number of jobs leaving the $i$-th queue at the current epoch,
$x_i$ is the number of jobs from the $i$-th queue for which no acknowledgements have been previously observed by the load-balancer,
$p_i$ is the probability that an acknowledgement is received by the load-balancer in the current epoch and
$L_i$ is the distribution from which $l_i$ is sampled for Eq. \eqref{eq:augmented_state}.

We have chosen the binomial distribution because it generally captures the fact that only a subset of the sent, $\min( b_i,  k_i) +  x_i$, acknowledgements are successfully observed at the load-balancer in the current epoch.
At the cost of simplifying the usually correlated delays of jobs, this model helps to obtain tractable results.
For example, $p=0.6$ would mean that out of all the acknowledgements sent, only $60$\% are expected to be received by the load-balancer in that epoch while $40$\% are expected to be delayed.
Similarly, $p_i=1$ would then represent the case of no delay.
Note that any other distribution describing the delay model can be numerically evaluated.
The number of acknowledgements from all servers that are not observed in this epoch are accounted for in the next epoch in $\mathbf x^\prime$.
The previously introduced observations $\mathbf{z}$ are essentially the received acknowledgements, i.e., $\mathbf{z} =\mathbf{y}'=\mathbf l$.
Since only the vector $\mathbf{y}'$ of the state $\mathbf s'$ (Eq. \eqref{eq:augmented_state}) is observed by POL, a partial observability is established.
We note that one limitation of this model is due to the delay independence assumption that lies below the used Binomial distribution.
\subsection{Job Acknowledgement Distribution}
\label{subsection:job_acks}
Next, we discuss how to quantify the distribution of the number of jobs $k_i$  that can be served at the $i$-th queue in one inter-arrival time. The marginal probability
\begin{equation}
f_{K_i}(k_i)=\int_{0}^\infty f_{K_i \mid U}(k_i \mid u) f_U(u)  d u,
\label{eq:integral}
\end{equation}
 can be computed by noting \cite{ross2014introduction} 
\begin{equation}
f_{K_i \mid U}(k_i \mid u)=\mathbb{P}\left(\bar{V}^{(k_i)}_i \leq u\right) - \mathbb{P}\left(\bar{V}^{(k_i+1)}_i  \leq u \right),
\end{equation}
with $\bar{V}^{(k_i)}_i=\sum_{m=1}^{k_i} V^{(m)}_i$.
For the POMDP model, with exponentially distributed inter-arrival times $U \sim \ExpDis(\lambda)$,
with rate parameter $\lambda$,
and exponential service times for all servers
$V_i \sim \ExpDis(\mu_i)$,
with rate parameter $\mu_i$, the distribution $f_{K_i}(k_i)$ can be calculated in closed form. Since the service process corresponds to a Poisson process, we find the conditional distribution $f_{K_i \mid U}(k_i \mid u)$ as $K_i \mid u \sim \PoisDis( \mu_i u)$.
Carrying out the integral, in Eq. \eqref{eq:integral}, we find the number of jobs that can be served in one inter-arrival time follow a Geometric distribution $K_i \sim \GeomDis( \frac{\lambda }{\mu_i + \lambda} ),$
with $\frac{\lambda }{\mu_i + \lambda}$ denoting success probability, \cite{ross2014introduction}. 

For the POSMDP model with general inter-arrival and service time distributions closed form expressions for the marginal probability above are often not available, since, this would require closed form expressions for a $k$-fold convolution of the probability density function (pdf) of the service times.
Note that some corresponding general expressions exist as Laplace transforms where the difficulty is passed down to calculating the inverse transform.
Therefore, in such cases we will resort to a sampling based scheme for the marginal distribution $f_{K_i}(k_i)$, i.e.,
\begin{equation}
\begin{split}
&U \sim f_U(u) \\
&V^{(m)}_i \sim f_{V_i}(v_i), \quad m=1, 2, \dots\\
&\mathcal K_i = \left\lbrace j \in \mathbb N_0: \sum_{m=1}^{j} V^{(m)}_i \leq U \right\rbrace \\
&K_i=\max(\mathcal K_i).
\end{split}
\label{eq:sampling}
\end{equation}
In this numerical solution, we draw a random inter-arrival time $U$ and count the number $K_i$ of service times that fit in the inter-arrival interval.
Note that the number of jobs $K_i^{(t)}$ at time $t$  is not necessarily independent for the number of jobs $K_i^{(t+1)}$ in the next interval. The impact of this effect can be well demonstrated when the service time distribution is, e.g.,  heavy tailed.
Also note that the exact modelling of this behaviour would, in general, require an extended state space incorporating this memory effect.
Therefore, the sampling scheme can be seen as an approximation to the exact system behaviour.
\subsection{Inferring Arrivals and System Parameters}
\label{sec:inference}
For POL to be deployed in an unknown environment, we may require an estimate of the inter-arrival and/or service rate densities.
For this purpose, we will resort to a Bayesian estimation approach to infer the densities $f_U(u)$ and $f_{V_i}(v_i)$.
We select a likelihood model $f(\mathcal{D} \mid \theta)$ for the data generation process and a prior $f(\theta)$, with model parameters $\Theta$. We assume we have access to data
$\mathcal{D}=\{d^{(1)},d^{(2)}, \dots, d^{(n)} \}$,
where  $d^{(j)}$ is inter-arrival times between the $j$-th and $j+1$-st job arrival event that is observed by the load-balancer or the service times for each server.
For inference-based load balancing in POL we use the inferred distribution of the inter-arrival times as in the posterior predictive $f(d^\ast \mid \mathcal{D})=\int f(\theta \mid \mathcal{D}) f(d^\ast \mid \theta) d \theta$,
of a new data point~$d^\ast$, which is then used in the sampling simulator, see Eq.~\eqref{eq:sampling}.
The same can be done with the data for the service times.
We will now describe some models of different complexities for the data generation. Since, most models do not admit a closed form solution, we resort to a Monte Carlo sampling approach to sample from the posterior predictive.

\subsubsection{Inference for exponential inter-arrival times}
\label{sec:exponential}
Here, we briefly show the calculation for the posterior distribution and posterior predictive distribution for renewal job arrivals with exponentially distributed inter-arrival times. For the likelihood model we assume
\begin{equation*}
D^{(j)} \mid m \sim  \ExpDis(m),\quad j=1, \dots, n
\end{equation*}
where $m$ is the rate parameter of the exponential distribution. We use a conjugate Gamma prior $M \sim \GamDis(\alpha_0, \beta_0)$.
Hence, the posterior distribution is
\begin{equation*}
M \mid \mathcal{D} \sim \GamDis(\alpha_0 +n, \beta_0+ \sum_{j=1}^n d^{(j)})
\end{equation*}
And the posterior predictive distribution is found as
\begin{equation*}
D^\ast \mid \mathcal{D} \sim  \TPDis(\alpha_0 +n, \beta_0+ \sum_{j=1}^n d^{(j)}),
\end{equation*}
where  $\TPDis(\alpha, \beta)$ denotes the translated Pareto distribution.
\subsubsection{Inference for Gamma distributed inter-arrival times}
\label{sec:gamma_likelihood}
In case of a gamma likelihood of the form
\begin{equation*}
D^{(j)} \mid \alpha, \beta  \sim  \GamDis(\alpha, \beta),\quad j=1, \dots, n
\end{equation*}
we use independent Gamma priors for the shape and the rate, with $A \sim \GamDis( \alpha_0, \beta_0)$ and $B \sim\GamDis(\alpha_1, \beta_1)$.
Finally, we sample from the  posterior predictive using Hamiltonian Monte Carlo (HMC)~\cite{hoffman2014no}, which can be implemented using a probabilistic programming language, e.g. using PyMC3 \cite{salvatier2016probabilistic}.

\subsubsection{Service times distributed as an infinite Gamma mixture}
\label{sec:mixture}
Here, we present a framework to non-parametrically infer the posterior distribution.
We use an approximate Dirichlet process mixture model which
can be regarded as an infinite mixture model~\cite{ghosal2017fundamentals}.
We use a gamma distribution for the observation model
\begin{equation} \phi(d \mid \theta_i) \propto d^{a_i-1} e^{-b_i d},
\end{equation}
with mixture parameters $a_i$ and $b_i$. For the base measure $G_0$, i.e., the prior distribution of the mixture parameters, we use $G_0=F_{A_i} \times F_{B_i}$, with $A_i \sim \GamDis( 1,1)$ and $B_i \sim \GamDis(1,1)$.
The  truncated stick-breaking approximation is then given by
\begin{equation*}
\begin{split}
&M \sim \GamDis(1,1), \quad B_i \mid m  \sim  \BetaDis(1, m), \quad i=1,\dots, c-1\\
&W_i = \beta_i \prod_{j=i-1}^i (1-B_j), \quad i=1,\dots, c-1\\
& W_c =1 -\sum_{j=1}^{c-1} W_j, \quad \Theta_i \sim G_0\\
&D^{(j)} \mid w_1,\dots, w_c, \theta_1, \dots, \theta_c   \sim   \sum_{i=1}^c w_i \phi(d \mid \theta_i), \; j=1, \dots, n,\\
\end{split}
\end{equation*}
which corresponds to a mixture of Gamma pdfs.
Here too, samples from the posterior predictive can be efficiently generated using HMC.
For the truncation point, the number of components $c$ in the formula above can be assessed using
\begin{equation}
c= \lceil2-\E[M] \log(\epsilon)\rceil=\lceil2-\log(\epsilon)\rceil,
\end{equation}
where $\epsilon$ is an upper bound on the total variation distance between the exact and truncated approximation. For example, we can choose $\epsilon=10^{-12}$, which corresponds to $c=30$ components.

\begin{algorithm*}
\caption{POL load-balancer with delayed feedback}
\label{alg}
  \begin{algorithmic}[1]
\State  \textbf{Input}:~$N, \lambda, \mu_1 \ldots \mu_N, \mathcal G, \eta,  \bm{\rho}^{(0)}, \mathcal R, \mathbf s_0, \kappa, b, T_m, T_e, Q_s$
  \State \textbf{Output}: $R_{\text{avg}}$, average reward for each time step $T_e$
  \State Initialize $R_m$
  \For {$t = 0, 1, \ldots, T_m$}
    \State Initialize $\mathbf u, \mathbf v^{(1)} \ldots \mathbf v^{(N)}, \text{tree } \bm{\Psi_0}, R_e, Q_s$ \Comment{$Q_s$ is the real world representation of the queuieng network.}
        \For {$t = 0, 1, \ldots, T_e$}
            \State $\bm{\Psi_{t+1}}$ = SimulateTree($\bm{\Psi_t}, \mathcal G$)  \Comment{See the pseudocode given in \cite{silver2010monte}}
            \State $a_{t+1} \rightarrow \argmax_a \mathcal R(\mathbf s, a) $
            \State $o_{t+1}, s_{t+1}, r_{t+1}$ = $Q_s(a_{t+1})$     
            \State $R_e \rightarrow R_e  \cup r_{t+1}$  \Comment{Collect reward for each epoch}
                \State $\bm{\rho}^{(t+1)}$ = UpdateBeliefandTree($\bm{\Psi_{t+1}}, o_{t+1}, a_{t+1}, \mathcal G$)
        \EndFor
        \State $R_m \rightarrow R_m  + R_e$ 
   \EndFor 
   \State $R_{\text{avg}} = \frac{R_m}{T_m}$
   \State  \textbf{Return} $R_{\text{avg}} $
        \Function{UpdateBeliefandTree}{$\bm{\Psi}, o, a, \mathcal G$}
    \State Initialize $K_s=\{\}, W_s = =\{\} $ \Comment{Set of particles and their weights}
    \Repeat
      \State ~$s', o', r'~\sim~\mathcal G(s, a)$, \quad $s \sim \bm \Psi(\text{root)}$
      \State $w_s = p(s' | o', a)$
      \State $K_s \rightarrow K_s \cup s', W_s \rightarrow W_s \cup w_s$
        \Until Timeout()
        \State Generate particles from $K_s$ according to weights $W_s$ \Comment{SIR particle filter}
      \State Update root and prune tree $\bm{\Psi}$
    \EndFunction
  \end{algorithmic}

\end{algorithm*}

\subsection{Reward Function Design}
\label{subsec:reward_functions}
A main difference of our approach to \emph{explicitly} defining a load balancing algorithm is that we provide the algorithm designer with the flexibility to set different optimization objectives for the load-balancer and correspondingly obtain the optimal policy by solving the \pomdp or the \posmdp. This is carried out through the design of the reward function $\mathcal R$ as defined in Section~\ref{subsec:primer_pomdp}. The optimal policy $\pi^*$ maximizes the expected discounted reward as $\pi^* =  \argmax_{\pi} \sum^{\infty}_{t=0}  \E_{\pi}[\gamma^t R^{(t)}]$ where $R^{(t)} = \mathcal{R}(S^{(t+1)},S^{(t)},A^{(t)})$, with $\gamma<1$.
In the following, we discuss several reward functions $\mathcal R$ in the context of mapping incoming jobs to the parallel finite queues, see Fig.~\ref{fig:queuein_dynamics}.

\noindent\textbf{Minimize queue lengths:} A reward function which aims to minimize the overall number of jobs waiting in the system. This objective that can be formalized as
\begin{equation}
    \mathcal R(\mathbf{s}' , \mathbf{s}, a)=-\sum_{i=1}^N b_i'
  \label{eq:linear}
\end{equation}
as it takes the sum of all queue fillings. Similarly, a polynomial or an \textit{exponential} reward function, such as
\begin{equation}
\mathcal R(\mathbf{s}' , \mathbf{s}, a)=-\sum_{i=1}^N \chi^{b_i'}
\label{eq:expon}
\end{equation}
For a fixed overall number of jobs in the system and $\chi > 1$, this objective tends to balance queue lengths, e.g., if total jobs in the system are $10$ and $N=2$ then an allocation of $[5,5]$ jobs will have much higher reward than $[9,1]$ allocation.
Using the \textit{variance} amongst the current queue fillings also balances the load on queues. The reward function is then given as:
\begin{equation}
 \mathcal R(\mathbf{s}' , \mathbf{s}, a)=\mathrm{Var}(b_i,\dots,b_n)
   \label{eq:variance}
  \vspace{-5pt}
\end{equation}

Note, however, that balancing queue lengths does not necessarily lead to lower delays if the servers are heterogeneous.
Hence, \textit{proportional allocation}
provides more reward when jobs are mapped to the faster server as
\begin{equation}
    \mathcal R(\mathbf{s}' , \mathbf{s}, a)=-\sum_{i=1}^N \frac{b_i'}{\mu_i}
\label{eq:prop}
\end{equation}

\noindent\textbf{Minimize loss events:}
To prevent job losses, we can also formulate a reward function that penalizes actions that lead to fully filled queues, i.e.,
 \begin{equation}
  \mathcal R(\mathbf{s}' , \mathbf{s}, a)=-\sum_{i=1}^N\mathbbm{1}(b_i'=\bar{b}_i)
  \label{eq:pkt_drop}
  \vspace{-5pt}
\end{equation}
The indicator function evaluates to one only when the corresponding queue is full.

\noindent\textbf{Minimize idle events:} One might also require that the parallel system remains work-conserving, i.e., no server is idling, as this essentially wastes capacity. Hence, in the simplest case we can formulate a reward function of the form 
 \begin{equation}
 \mathcal R(\mathbf{s}' , \mathbf{s}, a)=- \sum_i \mathbbm{1}(b_i'=0)
   \label{eq:idle}
  \vspace{-5pt}
 \end{equation}

Note that some of the reward functions above can be combined, e.g., in a weighted form such as the following.

\begin{IEEEeqnarray}{rCl}
  \mathcal R(\mathbf{s}', \mathbf{s}, a)=-  \left[\sum_{i=1}^N b_i' + \kappa \mathbbm{1}(b_i'=\bar{b}_i) \right],
  \label{eq:linear_reward_function}
\end{IEEEeqnarray}

where, $b_i'$ is the buffer state after taking action $a$ and the constant weight $\kappa > 0$ is used to scale the impact of the events of job drops to the impact of the buffer filling on the reward.

\subsection{Partial Observability Load-Balancer: A Monte Carlo Approach for Delayed Acknowledgements}
\label{subsec:model_simulations}

In this section, we outline our approach to solve the partially observable system for the job routing problem in parallel queuing systems with delayed acknowledgements.
Our solution is an alternate technique to Dynamic Programming and is based on a combination of the Monte Carlo Tree Search (MCTS) algorithm \cite{silver2010monte} and Sequential Importance Resampling (SIR) particle filter.

The reason for choosing an MCTS algorithm is that load balancing problems, like the one presented in this work, can span to very large state spaces. In these scenarios, solution methods based on dynamic programming \cite{bellman1966dynamic} often break due to the \emph{curse of dimensionality}.
MCTS solves this problem by using a sampling based heuristic approach to construct a search tree to represent different states of the system, the possible actions in those states and the expected value of taking each action.  
In recent years, these techniques have been shown to yield exceptional results in solving very large decision-making problems \cite{silver2016mastering},\cite{silver2018general}.

One main contribution of the work at hand lies in the design of a simulator\footnote{\url{https://github.com/AnamTahir7/Partially-Observable-Load-Balancer}}, 
$\mathcal G$, which incorporates the properties of the queuing model discussed above, into the algorithm. 
The simulator  $\mathcal G$, provides the next state $\mathbf s^{(t+1)}$, the observation $\mathbf z^{(t+1)}$ and the reward $R^{(t+1)}$, when given the current state $\mathbf s^t$ of the system and the taken action $a^t$ as input,
\begin{equation}
\mathbf s^{(t+1)}, \mathbf z^{(t+1)}, R^{(t+1)} \mid \mathbf s^t, a^t \sim \mathcal G(\mathbf{s}, a ).
\label{eq:simulator}
\end{equation}
This simulator $\mathcal G$, is used in the MCTS algorithm to rollout simulations of different possible trajectories in the search tree.
Each trajectory is a path in the search tree starting from the current belief state of the system and expanding (using $\mathcal G$) to a certain depth.  
While traversing through the search tree, the trajectories (actions) are chosen using the Upper Confidence bounds for Trees algorithm (UCT), which is an improvement over the greedy-action selection \cite{kocsis2006bandit}. 
In UCT the upper confidence bounds guide the selection of the next action by trading off between exploiting the actions with the highest expected reward up till now and exploring the actions with unknown rewards.
 
At every decision epoch POL starts with a certain belief on the state of the system, $\bm{\rho}^{(t)}$, see Section \ref{subsec:primer_pomdp} for a formal definition, which is represented with particles and also used as the root of the search tree. 
Starting from the root, i.e., the current belief, the search tree uses UCT to simulate the system for a given depth (fixed here to tree depth of 10), after which the action $a$ with has the highest expected reward is chosen.
The trajectories for all other actions are then pruned from the search tree since they are no longer possible. 
This is done to avoid letting the tree grow infinitely large.

\begin{figure*}
\center
\includegraphics[scale=.5]{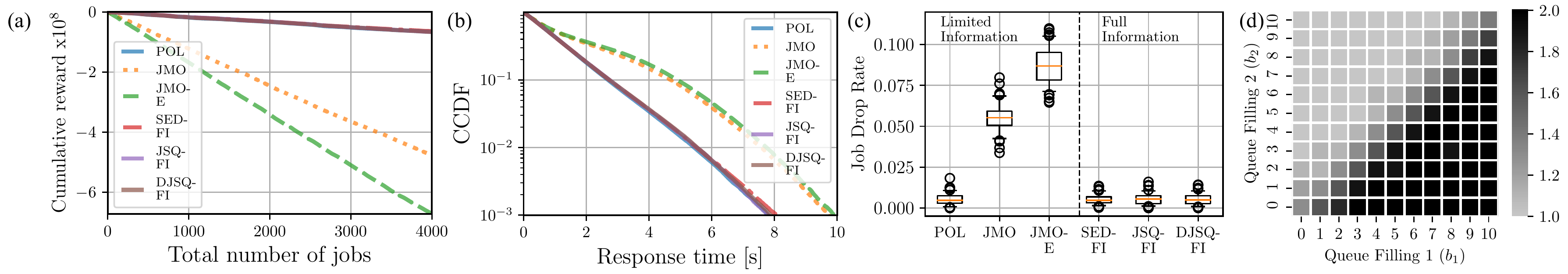}
        \caption{$N=2$ heterogeneous servers with exponential service rates $\mu_1=4$, $\mu_2=2$. Job inter arrival times are exponentially distributed with rate $\lambda=5$. POL outperforms the algorithms with limited information, JMO and JMO-E, in terms of (a) cumulative reward (higher is better), (b) response time (lower is better),  and (c) job drops (lower is better).
Although POL \emph{only observes} the randomly delayed job acknowledgements, while SED-FI, DJSQ-FI and JSQ-FI know the \textbf{exact buffer fillings and service times / rates}, POL still has comparable performance. The heatmap (d)
shows the allocation preference of POL based on how filled the queue is $b_1(b_2)$, i.e., the label bar $1.0$($2.0$) denotes the allocation to queue $1(2)$, respectively. }
        \label{fig:het2-exp-arr-exp-srv}
\end{figure*}


Once the action $a$ is taken and the job is allocated to a certain queue, the load-balancer receives real observations,~$\textbf{z}=\mathbf l$, from the system.
These observations are the randomly delayed acknowledgements from the servers.
The load-balancer then uses the received observations as an input to the SIR particle filter, in order to update its belief of the state the system is in now.
The weights given to each particle (state), $w(s_i) = p(z_i|s_i)$, while \textit{resampling} in this SIR particle filter were designed to incorporate our queuing system and its delay model, Eq. \eqref{eq:binomial}. 
After applying the SIR filter, we will have the new set of particles representing the current belief state of the system, $\bm{\rho}^{(t)}$.
These particles are then used to sample the states for simulating the search tree and finding the optimal action at the next epoch \footnote{Note that in POL receiving observations, action selection and belief update all happen at each decision epoch, which is why it is possible to model the system as a POSMDP \cite{yu2006approximate} as well.}. 
It is shown in Theorem $1$ of \cite{silver2010monte} for a POMDP and in Theorem $2$ of \cite{vien2015hierarchical} for a POSMDP, that the MCTS converges to an optimal policy.
 
To summarize, at every job arrival, POL simulates the tree from the root. The root contains the current set of belief particles. POL  then acts on the real environment using the action which maximizes the expected value at the root. On taking the action, POL gets a real observation. This observation is the acknowledgement that is subject to delays. Using this observation and an SIR particle filter, POL updates its set of particles (belief state) of the system for the next arrival. The pseudocode of the working of POL is given in Algorithm \ref{alg}.

\section{Simulation results}
\label{sec:sim_results}
In the following, we show numerical evaluation results for the proposed Partial Observability Load-Balancer (POL), under randomly delayed acknowledgements.
Recall that if the acknowledgement is not observed in the current inter-arrival time, it is not accumulated into the future observations.
In order to evaluate the impact of delayed observations, we consider in our simulations a probability of $p_i=0.6$ $\forall i$ in Eq. \eqref{eq:binomial}, if not stated otherwise.
This means that an acknowledgement is delayed until the next epochs with probability: $1-p_i=0.4$.
We set the buffer size, $b_i$, for all queues to $10$ jobs. 
This value for $b_i$ was chosen arbitrarily, and any other value can be used.
Further, if not explicitly given, we use the \textit{combined reward function} given in Eq. \eqref{eq:linear_reward_function} with $\kappa=100$, since we aim to  avoid job drops in the system.
We consider the system depicted in Fig.~\ref{fig:queuein_dynamics} for both cases of heterogeneous and homogeneous servers.
In particular, we show numerical results comparing POL to different variants of load balancing strategies (with and without full system information) with respect to:
\begin{itemize}
\item the log complementary cumulative distribution function (CCDF) of the empirical job response time (measured from the time a job enters the queue until it completes service and leaves, lower is better).\textit{ This is done only for the jobs which are not dropped},
\item the empirical distribution of the job drop rate (measured over all simulation runs where for each run we track the number of jobs dropped out of all jobs received per run, lower is better),
\item and the cumulative reward (higher is better).
\end{itemize}
 The evaluation box plots are based on~$T_m=100$ independent runs of~$T_e=5\cdot10^3$ jobs with whiskers at~$[0.5, 0.95]$ percentiles. 
For every independent run, a new set of inter-arrival and service times are sampled based on the chosen distributions. These sampled times are then used by all load-balancing policies in that run in order to do variance reduction, according to the Common Random Numbers (CRN) technique \cite{l1994efficiency}. The plotted results are an average of $100$ such Monte carlo simulations. 

The chosen inter-arrival and service time distributions are mentioned with the figures, with unit of measurement req/sec.
The offered load ratio $(\eta:=\lambda /\sum_i\mu_i )$ is used to describe the ratio between arrival rate and the combined service rate. 
The higher the value of $\eta$, the higher is the job load on the system.

\subsection{Overview of compared Load-Balancers}
We compare POL to the following load balancing strategies:
\subsubsection{Full information (FI) strategies} These strategies have access to the exact buffer length of queues at the time of each job arrival, and also know the arrival rate and the service rates of the servers.
\begin{itemize}
\item JSQ-FI: Join-the-Shortest-Queue assigns the incoming job to the server with the smallest buffer filling.
\item DJSQ-FI: Join the shortest out of $d$ randomly selected queues. If not stated otherwise, $d=2$ has been used for our experiments.
\item SED-FI:  Shortest-Expected-Delay assigns the incoming job to the server with the minimum fraction of the current buffer filling divided by the average service rate.
\end{itemize}
\subsubsection{Limited information (LI) strategies} These strategies, similar to POL, only have access to the randomly delayed acknowledgements.
\begin{itemize}
\item JMO: Join-the-Most-Observations maps an incoming job to the server that has generated the most observations, i.e., received acknowledgements, in the last inter-arrival epoch. This might lead to servers becoming and remaining idle (stale).
\item JMO-E (with Exploration): with probability $0.2$ randomly chooses an idle server and with probability $0.8$ performs JMO.
\end{itemize}
For all strategies, ties are broken randomly. 


\begin{figure}
\center
\includegraphics[scale=.5]{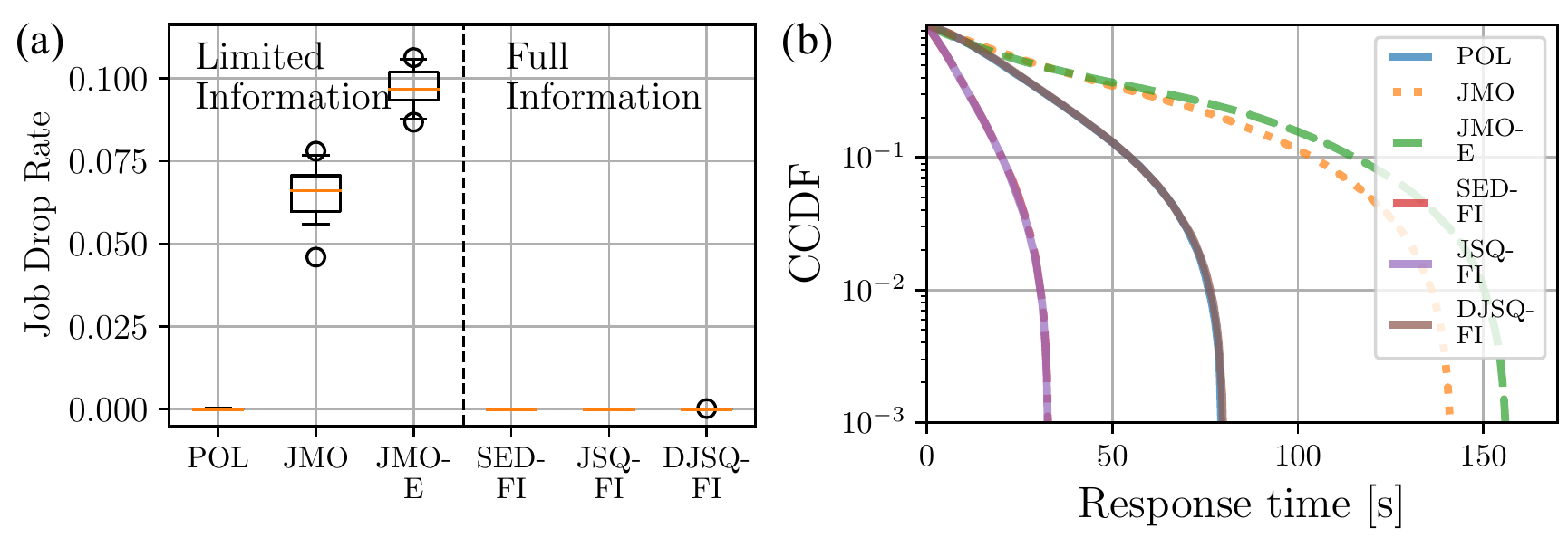}
        \caption{$N=50$ heterogeneous servers with job inter-arrival times and service times described by an exponential distribution, with the offered load $(\eta \approx 1)$. POL outperforms the other algorithms with limited information, JMO and JMO-E, in terms of both (a) job drops and (b) response time. And has comparable performance to the full information strategies, SED-FI, DJSQ-FI and JSQ-FI.}
\label{fig:het50-exp-arr-exp-srv}
\end{figure}


\begin{figure}
\center
\includegraphics[scale=.5]{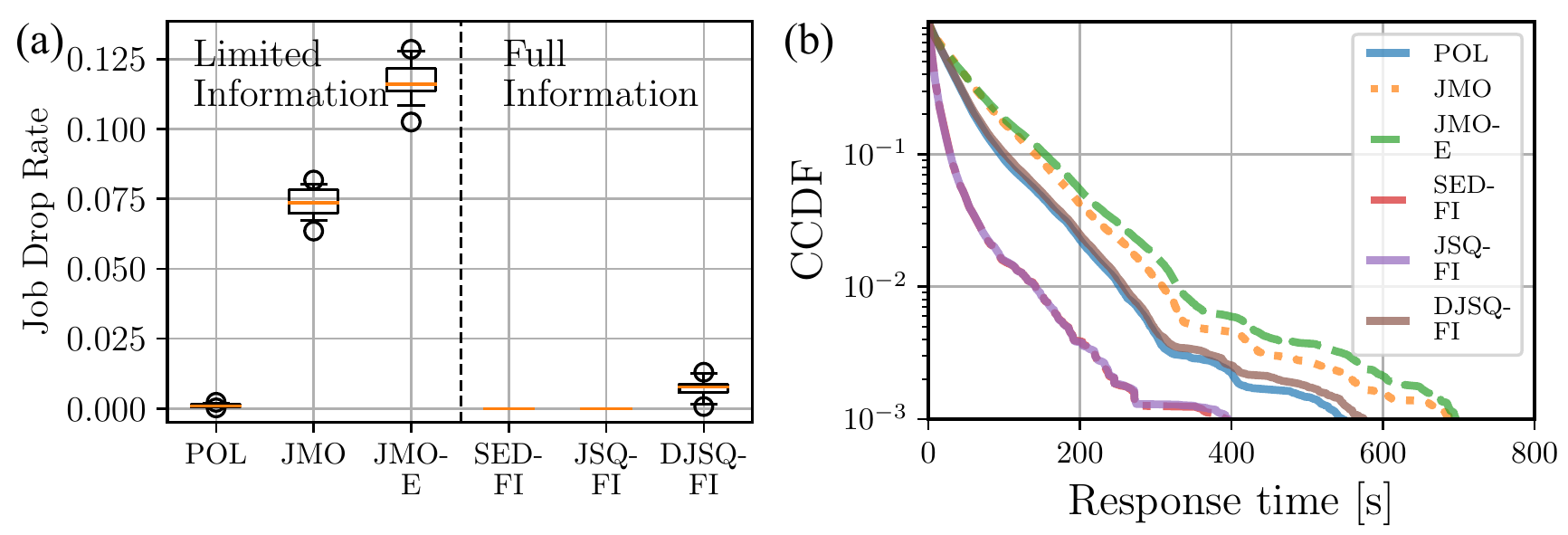}
         \caption{$N=50$ heterogeneous servers with gamma arrivals and Pareto service times, with offered load $(\eta \approx 1)$.  
         The effect of the heavy tailed Pareto distribution can be seen in the response plot (b).
          In terms of job drops, POL outperforms limited information (LI) strategies as well as FI strategy, DJSQ-FI. 
          }
 \label{fig:het50-gamma-arr-par-srv}
\end{figure}

    

\begin{figure}
\center
\includegraphics[scale=0.5]{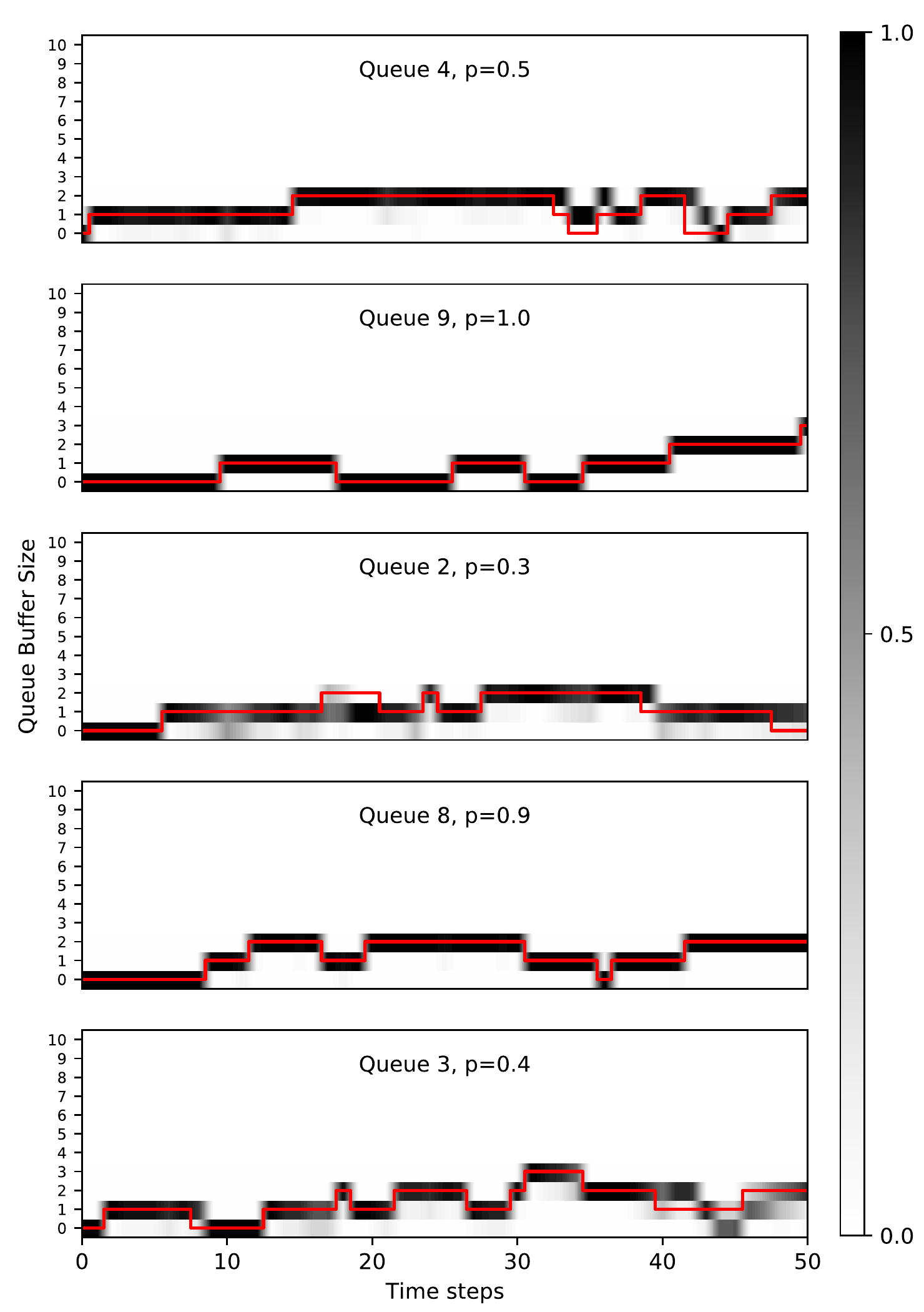}
    \caption{$N=10$ homogenous servers from which $5$ were chosen at random to see their sample run for $50$ time steps. The delay $p$ for the acknowledgements from each of the server was also allocated randomly, ranging from $0.1$ to $1.0$.  In each subplot is given the queue number and its delay, $p$. The solid red line trajectory is the true sample path for each queue (\emph{not known to POL}), while the shaded region around it is the belief probability that POL has for each state at each time step.   
Exponentially distributed inter-arrival and service times were used, with the offered load $\eta \approx 1$. 
POL, with the help of SIR particle filter, is able to track the real state of the system, as long as the delay is not too high, which is why it is able to perform as good as FI strategies.}
 \label{fig:sample run}
\end{figure}

    
\subsection{Numerical Results}
We first consider a system with $N=2$ heterogeneous servers with exponentially distributed service times with rates $\mu_1=4$ and $\mu_2=2$. 
The inter-arrival times are also exponentially distributed with rate $\lambda =5$.
Fig.~\ref{fig:het2-exp-arr-exp-srv}  shows the numerical comparison of POL with other load-balancers. 
Observe that, even though POL does not have access to the exact state of parallel systems and also the acknowledgements from the different systems are randomly delayed, it still achieves comparable results to \textit{full information strategies}, while it outperforms the other \textit{limited information strategies}. 
This is because in the other \textit{limited information strategies}, the initially chosen queues play a key role. Since the queue to which more jobs are sent, will also give back more observations (acknowledgements), and JMO and JMO-E will keep sending to those queues, resulting in job drops.
The overlap in response time indicates the similarity of the policies of the strategies, especially for high offered load when most of the finite queues will be full.
The heatmap in, Fig. \ref{fig:het2-exp-arr-exp-srv}(d), represents the policy of POL at each buffer filling state.
It can be seen that higher priority is given to the faster server, $\mu_1$, having buffer filling $b_1$. 
The light(dark) regions in the heatmap corresponds to the state where jobs are allocated to server $1$($2$).
\textit{This heatmap shows that for very possible state of the two queues $s=\{b_1,b_2\}$, even with limited and delayed information, POL is able to allocate more jobs to the queue with lower filling or faster servers, similar to JSQ and SED.} Hence, it is able to perform almost as good as the FI strategies.

Fig. \ref{fig:het50-exp-arr-exp-srv} shows the performance of POL for $N=50$ heterogeneous servers.
The service and arrivals rates of this setup are kept that the offered load is $(\eta \approx 1)$.
This experiment shows that our load-balancer POL is scalable to perform well for large number of queues.
Next, we remain with the case of $N=50$ heterogeneous servers, however with inter-arrival times that are gamma distributed while the service times follow a heavy tailed Pareto distribution, with the offered load $(\eta \approx 1)$.
Fig. ~\ref{fig:het50-gamma-arr-par-srv} shows that here too, POL is able to outperform both the LI strategies and the FI strategies, DJSQ-F.
The other two FI strategies have better performance because they always have timely and exact information of the queues, which is unrealistic. Note that as we consider heavy-tailed distributions in this example the prediction of job acknowledgments by POL suffers, because of reasons discussed at the end of subsection~\ref{subsection:job_acks}.


\begin{figure}
\center
\includegraphics[scale=0.45]{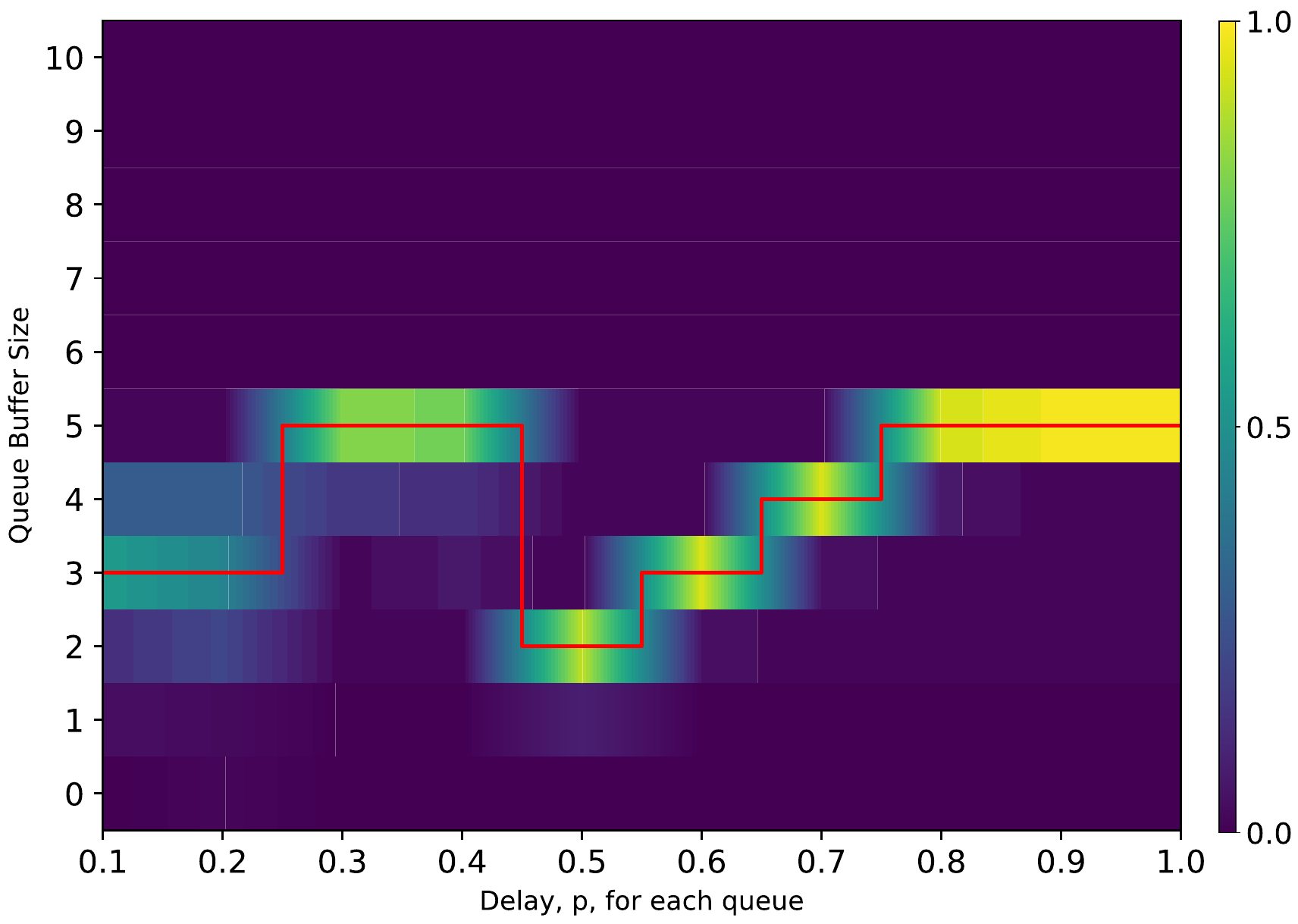}
    \caption{Visualization of belief state, based on the particles, of $N=10$ queues after $1000$ epochs. The x-axis gives the delay probability of each queue, ranging from $p=0.1$ (worse delay) to $p=1.0$ (no delay). The solid red line trajectory is the true state of each queue (not known to POL), while the shaded region around it is the belief probability that POL has for each queues' state after $1000$ time steps. It can be seen that as the acknowledgements become less delayed (going from $p=0.1$ to $p=1.0$)), the belief of POL gets closer to the true state of the queue. Exponentially distributed inter-arrival and service times were used, with the offered load, $\eta \approx 1$.}
 \label{fig:final run}
\end{figure}


\subsection{Sensitivity Analysis}

\begin{figure}
\center
\includegraphics[scale=0.5]{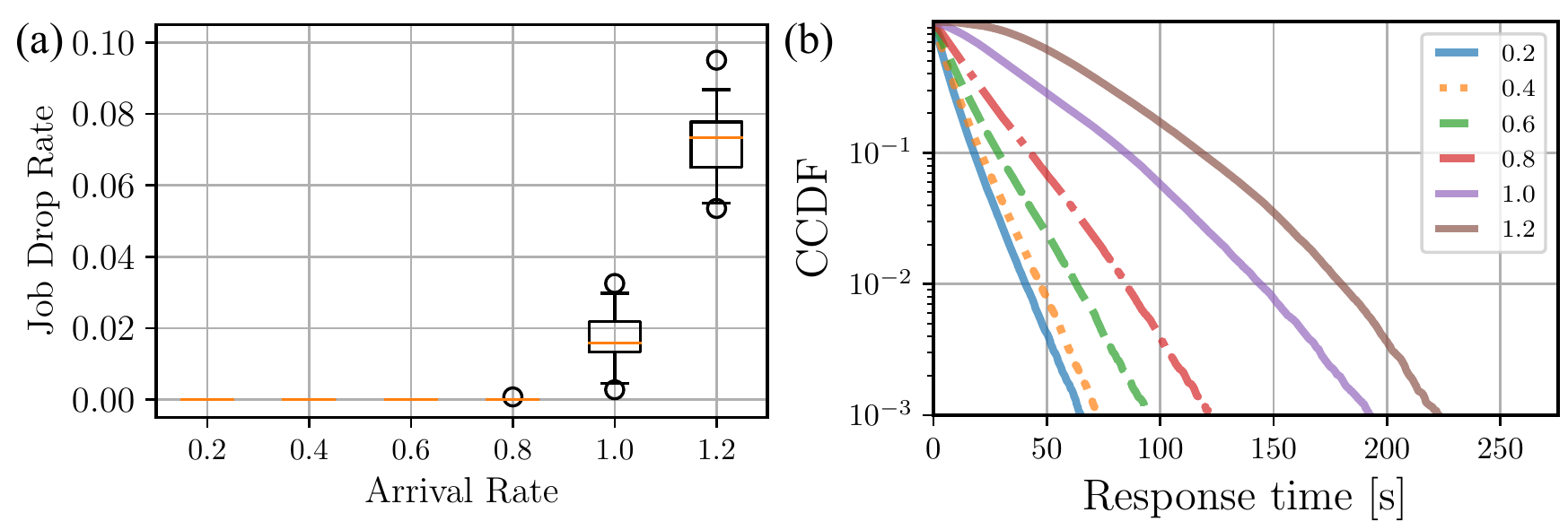}
    \caption{Varying offered load for a setup as in Fig.~\ref{fig:het50-exp-arr-exp-srv}. As long as the offered load is $\eta < 1$, POL has no job losses. For loads $\eta \ge 1$ we observe a load dependent exponential tail of the response time distribution.}
 \label{fig:vary arr rate}
\end{figure}



\begin{figure}
\center
\includegraphics[scale=0.5]{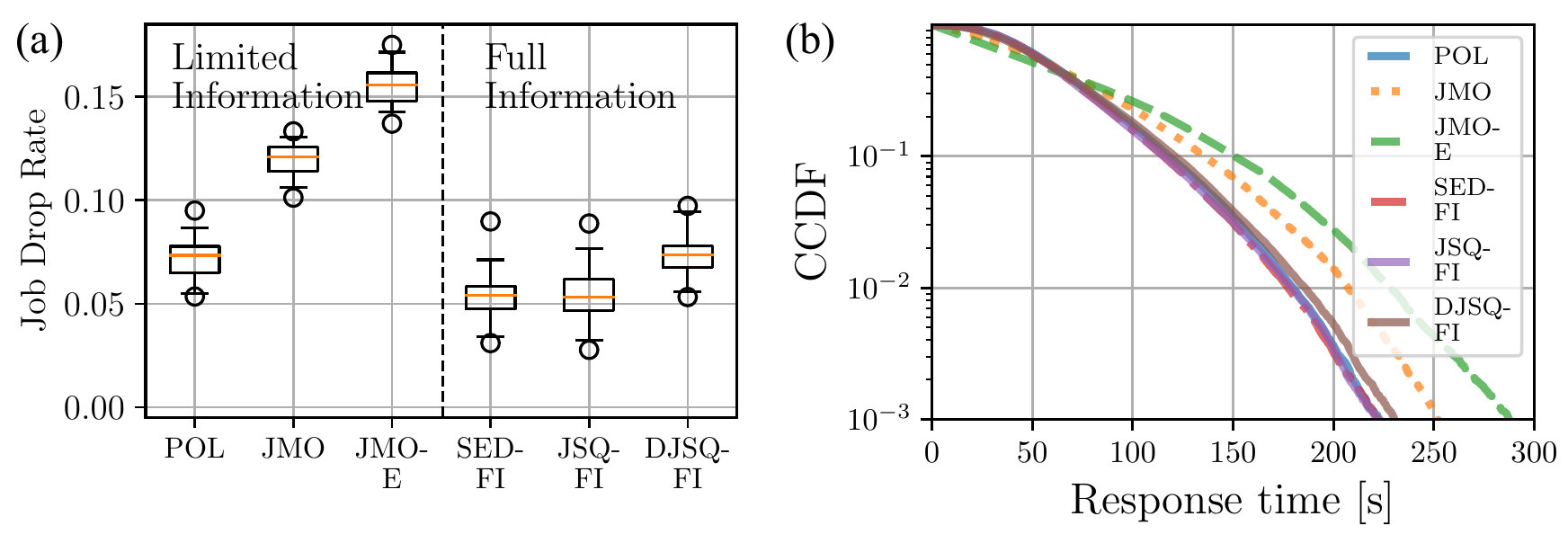}
    \caption{For  the setup from Fig. \ref{fig:vary arr rate} with  an offered load $\eta =1.2$: POL shows a comparable performance to full information (FI) strategies, while outperforming other limited information (LI) ones. }
 \label{fig:high-load}
\end{figure}

Next, we discuss the impact of the limited observations on POL under different acknowledgement delays, $p_i$. 
Recall, that POL is not able to observe the buffer fillings, but rather receives the randomly delayed acknowledgements of the served jobs. 
These delayed acknowledgements are used by the SIR filter of POL to keeps its belief of the state of the environment updated. 
Fig. \ref{fig:sample run} depicts sample runs showing the actual evolution of the job queue states (red solid line) and the belief (in shaded region) that POL has on each queue state at each time step, under different acknowledgement delays. 
Observe that increasing delays (i.e., lower $p_i$) increases the uncertainty in belief of each state. 
However, POL is still able to track the system state for different delays for each server, which shows the efficiency of the SIR particle filter and also \textit{justifies the performance of POL to be as good as FI strategies}.
Having different delays in acknowledgements from each server reflects a distributed system, where network conditions may be different for each server and may lead to different delays in acknowledgements from different servers. 
Fig. \ref{fig:final run} visualizes the belief of POL on the state of each queue after $1000$ epochs, for different delays in acknowledgements in each queue.

In Fig. \ref{fig:vary arr rate} we analyse the performance of POL under varying offered loads ranging from $\eta=0.2$ to $\eta=1.2$.
It can be seen that POL has almost no job losses up to a load of $\eta = 1$.
Note the qualitative change of the response time distribution as the offered load reaches $\eta = 1$ and beyond. 
For lower offered load, the response time distribution resembles an exponentially tailed distribution which changes with $\eta = 1$ and beyond.
In Fig. \ref{fig:high-load} we show the performance comparison of different LI and FI strategies for the high offered load case of $\eta=1.2$.
This is done to show that even though POL has high job drops and response times, it outperforms the LI strategies and has comparable performance to the FI strategy, especially DJSQ-FI.


\begin{figure}
\center
\includegraphics[scale=0.49]{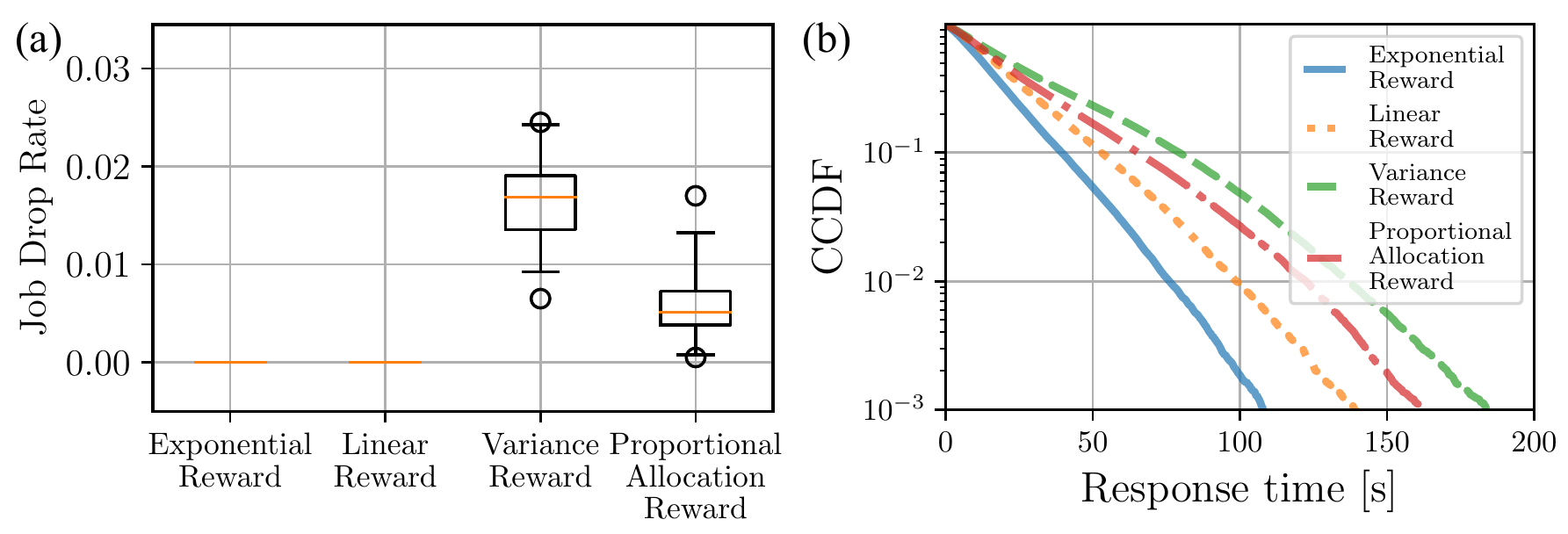}
    \caption{
    Comparison of different reward functions for $50$ homogeneous servers, with exponentially distributed inter-arrival and service times and offered load $\eta \approx 1$. }
 \label{fig:reward comp}
\end{figure}



\begin{figure}
\center
\includegraphics[scale=0.5]{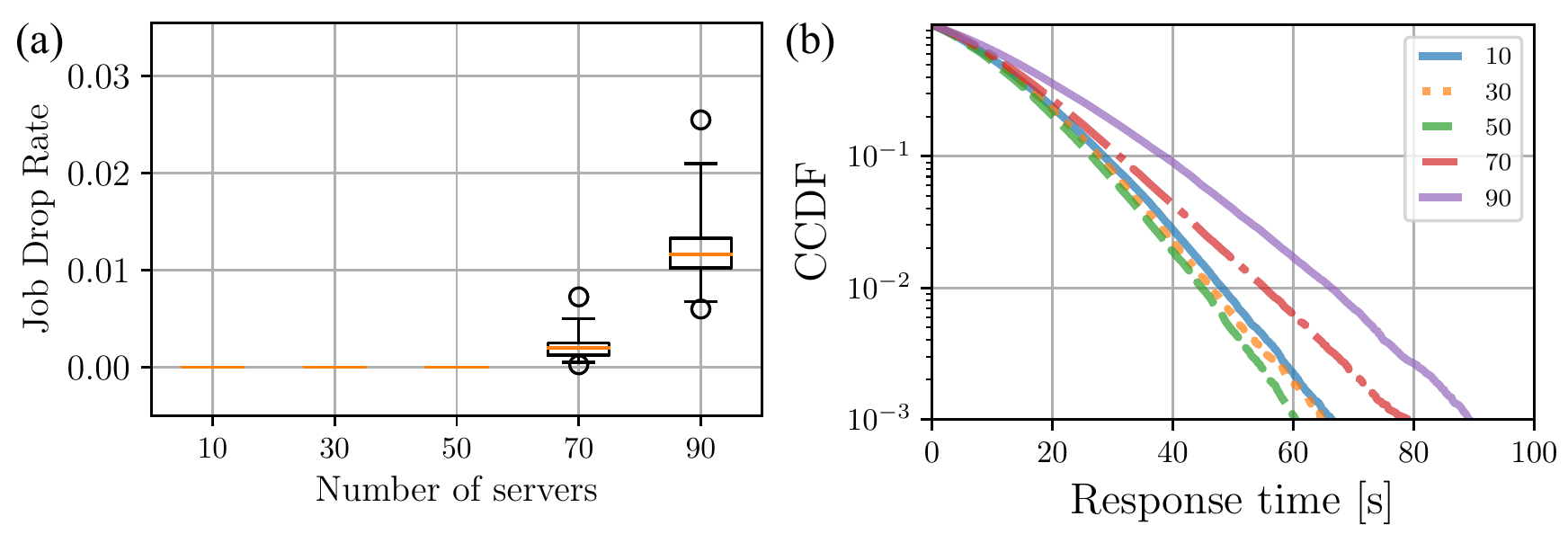}
         \caption{
         Homogeneous servers with exponentially distributed inter-arrival and service times. The number of servers are increased in intervals of $20$ servers, while keeping the offered load always $\eta = 0.99$. With a higher number of servers, less time is available for POL to simulate the tree and do belief update, resulting in a slight deterioration in the performance.}
 \label{fig:increasing-servers}
\end{figure}



\begin{figure}
\center
\includegraphics[scale=0.5]{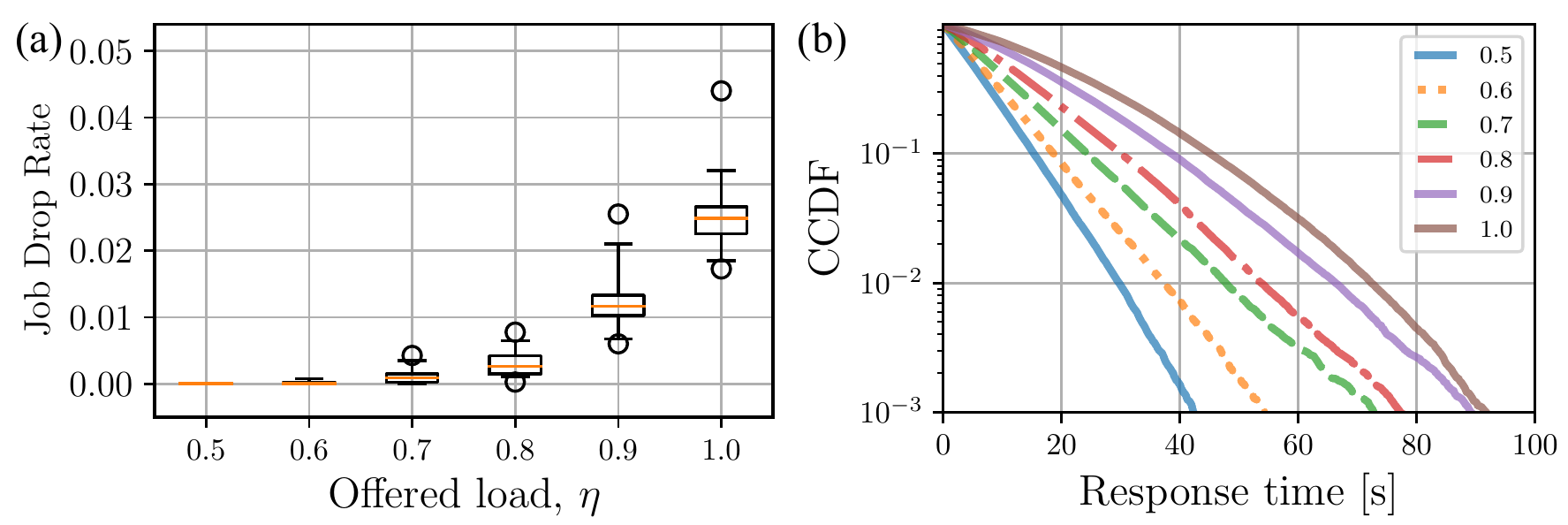}
         \caption{
         Performance of POL for $N=90$ homogeneous servers for different loads.}
 \label{fig:n90-load}
\end{figure}


For the sake of completeness, Fig. \ref{fig:reward comp} compares the performance of POL using different reward functions from subsection \ref{subsec:reward_functions}, while keeping all the other parameters the same.

\vspace{5pt}
\noindent \textit{Time Analysis of POL:}
POL consists of two main components: (i) Tree simulator for action evaluation and (ii) SIR particle filter for belief update. Both steps need to be done at every decision epoch, i.e. on every job arrival. Since we assume no queue at the load-balancer, POL needs to allocate the incoming job to one of the queues, before the next arrival.
Note that MCTS is a very successful online algorithm. Hence,  POL first takes a portion of the time between arrivals to simulate the tree and take a decision for the current arrival. Then takes that action on the real environment and based on the received delayed acknowledgement performs the belief update using the SIR particle filter, until the next job arrives.
As can be seen from the simulation results, POL is scalable in terms of number of servers and is able to handle high offered loads, $\eta$. Note that the code used to run POL here in the system simulation is the same that would be used in a deployment scenario. Next, we investigate the impact on the inter-arrival time between jobs on the load balancer performance as the inter-arrival time needs to be sufficient for POL to perform the above two steps at every job arrival. 

In Fig. \ref{fig:increasing-servers}, the number of homogeneous servers $N$ was increased from $10$ to $90$, while keeping the offered load fixed, i.e., $\eta=0.99$. The average service rate $\mu_i$ of the servers is kept fixed in all experiments, i.e. the increase in $N$ results in an increase in the sum of service rates, $\sum_i \mu_i$. 
Hence, to keep the offered load fixed with scale the job arrival rate accordingly. 
Firstly, this experiment demonstrates the scalability of POL in terms of number of servers. Although the state space of the system increases with the number of servers, hence, queues, POL manages to deal well with the increased state space. In POL we use MCTS adapted from POMCP \cite{silver2010monte}, so instead of considering the entire state space we have a fixed set of particles to represent the state based on our belief of the state, thus tapering the curse of dimensionality and space complexity.
As the inter-arrival time is the decision epoch we observe that the time given to POL to simulate the tree and do the belief update reduces, the effect of which can be seen as the slight decrease in performance as $N$ increases. 
It can be seen in Fig.~\ref{fig:n90-load}, that for $N=90$, lowering the load again, i.e. giving POL more time to decide, improves the performance of the system. This shows the trade-off between the load balancing performance, e.g. in terms of the response time and drop rate vs. the load which directly impacts the time provided to POL to make a decision. 

We believe this to be the current limitation of POL, however in future, step (i) can be further optimized using MCTS parallelization \cite{chaslot2008parallel}.
Note that the computational resources used also have a strong impact on the performance of POL. 
Here, we use a dedicated machine with an Intel(R) Xeon(R) CPU E5-2630 v2 @ 2.60GHz for all our experiments.

In the next section, we discuss the scenario when some system parameters are not known and need to be inferred from the available data.

\subsection{Experiments with trace data}
\label{sec:real_data}
\begin{figure}
\center
\includegraphics[scale=.54]{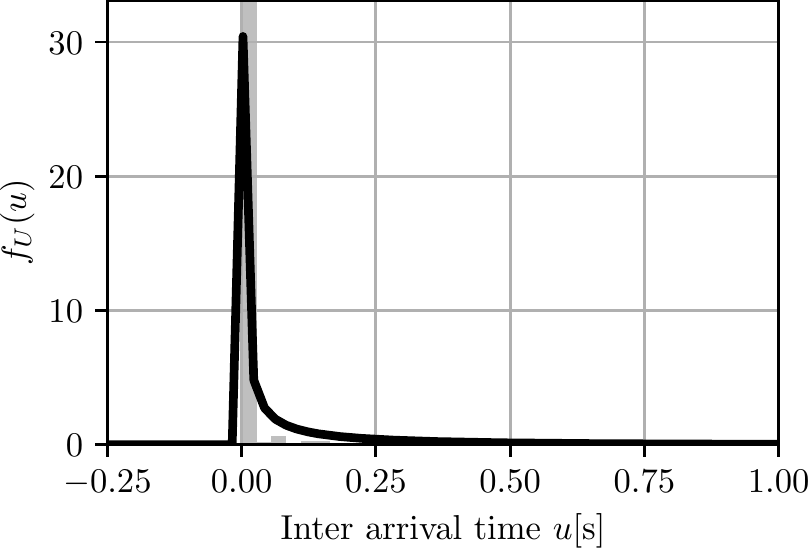}
 \caption{ The density estimate of the job inter-arrival times indicates that it is exponentially distributed.}
        \label{fig:arr}
\end{figure}

\begin{figure}
\center
\includegraphics[scale=.5]{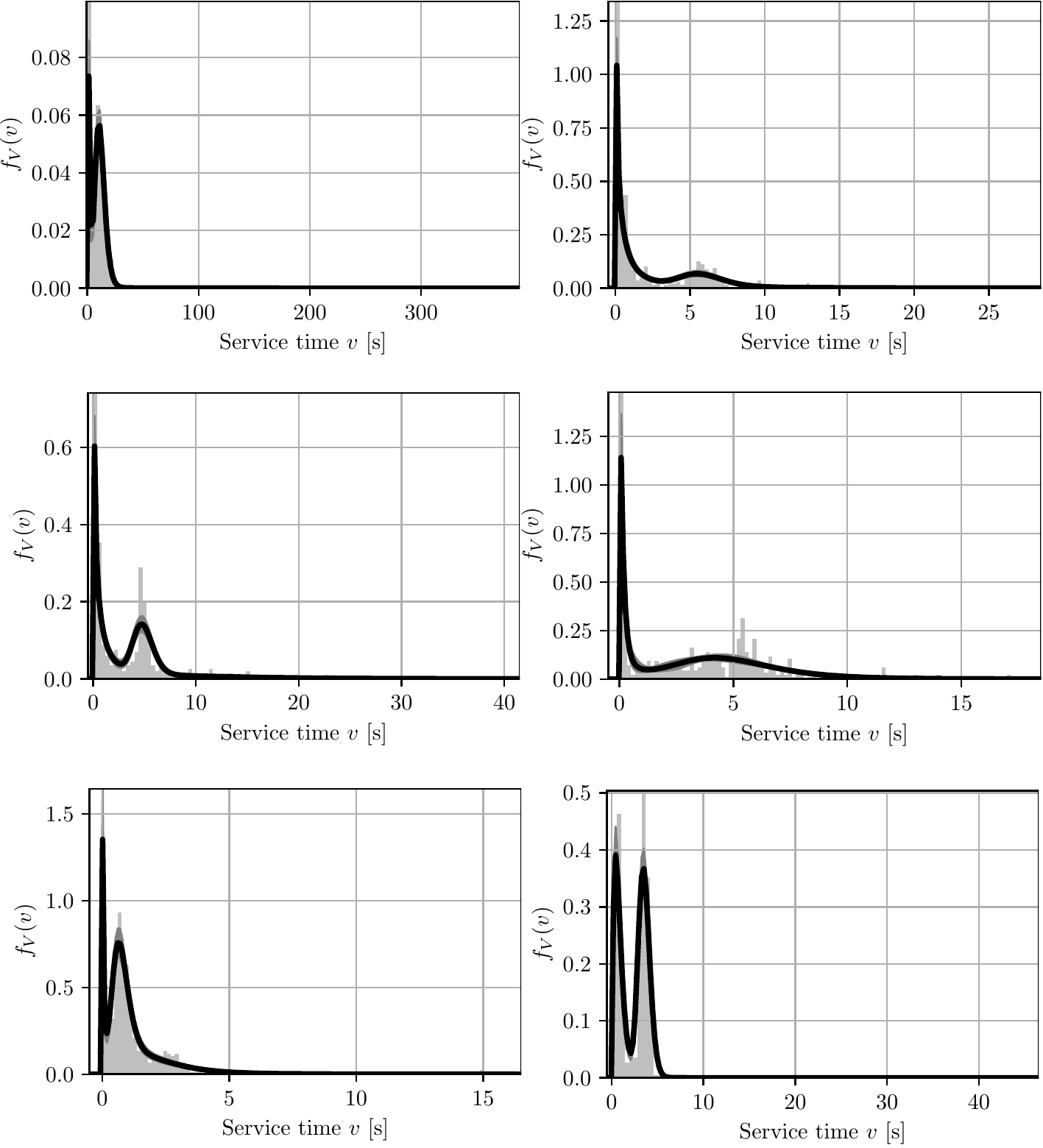}
 \caption{ The density estimate of the job service times using a Gamma Mixture Likelihood model indicates that the servers are heterogeneous with high service times.}
        \label{fig:srv}
\end{figure}

\begin{figure}
\center
\includegraphics[scale=.7]{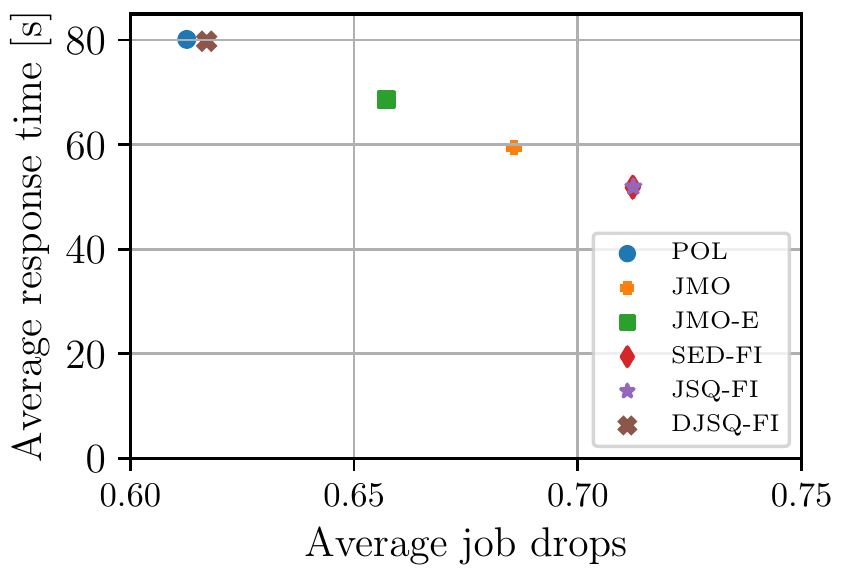}
 \caption{Tradeoff between response time and job drop: In terms of job drops, POL has comparable performance to DJSQ-FI, while it outperforms the other FI and all LI strategies. The additionally allocated jobs increase the overall average response time.}
        \label{fig:scatter}
\end{figure}

For the results of this section, we make use of \textit{Labeled Network Traffic Flow} data, provided by Kaggle in 2019 \cite{kaggle, kaggle2}.
We used the frameworks given in Subsection~\ref{sec:inference} to infer the underlying distributions of the inter-arrival and service times provided in this data set. 
The inferred distribution based on data of the inter-arrival times of the chosen source is given in Fig. \ref{fig:arr}, it can be seen to follow an exponential distribution with high arrival rate.
We then selected $20$ heterogeneous servers from the available data such that they all followed the Gamma Mixture distribution.
Gamma Mixture was selected to show the performance of POL with yet another type of service time distributions.
The empirical distribution as a histogram as well as the posterior mean estimate for some of these selected servers is given in Fig. \ref{fig:srv}. 
The hyperparameters used are: $c=3$, $a_i,b_i=1$, $m=1$.
POL makes use of the samples generated using the posterior predictions.

We assume that the servers have a finite buffer of size $B=10$ (arbitarily chosen) and a delay in acknowledgements of $p=0.6$.
Fig. \ref{fig:scatter} shows that even with limited information, POL has the lowest average job drops. However, due to the limited information available to POL and the reward function it is using, its average response time is as high as the full information strategy DJSQ-FI.
POL is able to allocate more jobs to the servers (due to fewer drops), which can come at a cost of higher response time for some jobs, which will be allocated to the slower servers.
Note that the reward function we used, \eqref{eq:linear_reward_function}, focuses on avoiding job drops and not on minimizing the response time.

\section{Discussion \& Conclusion }
\label{sec:conclusion}
In this work, we analyzed online algorithms for mapping incoming jobs to parallel and heterogeneous processing systems under partial observability constraints. This partial observability is rooted in the assumption that the entity controlling this mapping, denoted load-balancer, takes decisions only based on \emph{randomly delayed feedback} of the parallel systems. Unlike classical models that assume full knowledge of the parallel systems, e.g., knowing the queue lengths (Join-the-Shortest Queue  - JSQ) or additionally the job service times (Shortest Expected Delay - SED) this model is particularly suited for large distributed processing systems that only provide an acknowledgement-based feedback.

In addition to presenting a partially observable (semi-)Markov decision process model that captures the load balancing decisions in this parallel queuing system under delayed acknowledgements, we provide a {Partial Observable Load-Balancer (POL)} - to find near-optimal solutions online.
A particular strength of POL is that it allows to \emph{define the objectives of the system and lets it find the appropriate load balancing policy instead of manually defining a fixed one}.
It can also be used for any kinds of inter-arrival and service time distributions and is scalable to a large number of queues.
We numerically show that the POL load balancing policies obtained under partial observability are comparable to fixed policies such as JSQ, JSQ(d) and SED which have full information.
This is the case even though POL receives less, and in addition randomly delayed, informative feedback. 

One future direction is to parallelize the tree search part of POL to enable more trajectory simulations in a limited time. Use of GPUs can also be tested to investigate the system performance for a higher number of queues. 
Another direction for extending this work is to include the memory effect of the last served job in the model simulator, e.g., through state space extension.
Heterogeneous and batch jobs can also be considered. 
This work can also be extended to more than one load-balancer, working in parallel in a multi-agent manner.

\section*{Acknowledgment}
This work has been funded by the German Research Foundation (DFG) as part of sub-projects C3 and B4 within the Collaborative Research Center (CRC) 1053 – MAKI.

\bibliographystyle{IEEEtran}
\bibliography{bibliography_main}

\vspace{-1cm}
\begin{IEEEbiography}[{\includegraphics[width=1in,height=1.25in,clip,keepaspectratio]{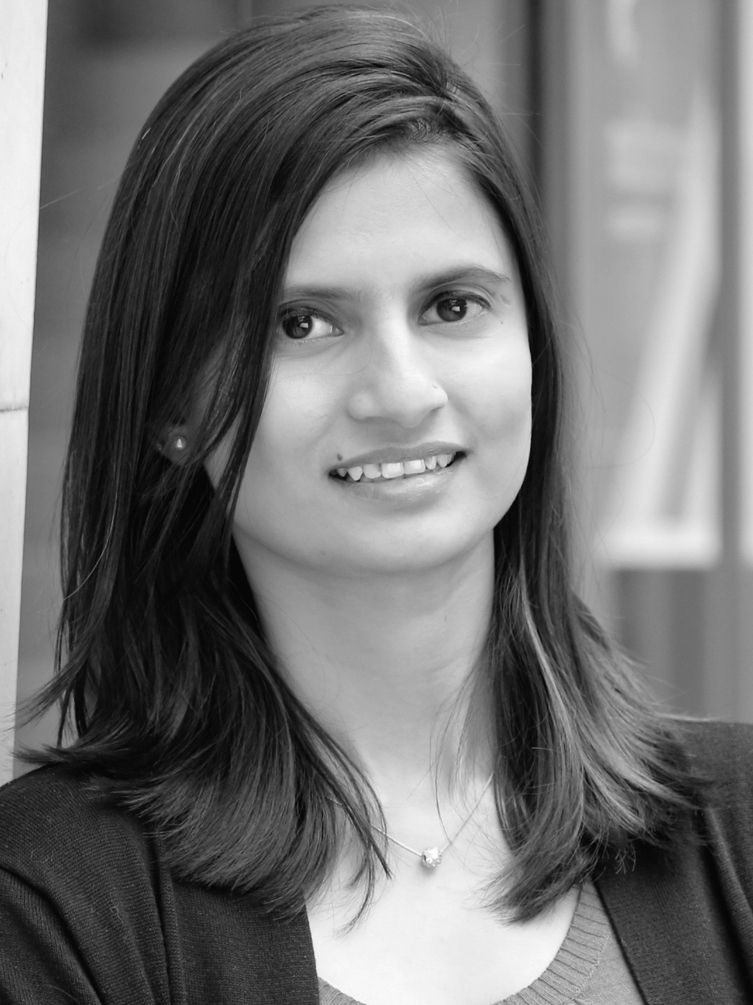}}]{Anam Tahir}
received her B.Sc. degree in electrical engineering and information technology from National University of Science and Technology, Pakistan. In 2018, she received her M.Sc. degree in electrical engineering and information technology from Technische Universität Darmstadt. Since November 2018, she is working as a research associate within the Self-Organizing Systems Lab at the Technische Universität Darmstadt. She is interested in planning and performance analysis of large queuing and other networked systems, particularly in uncertain environments.
\end{IEEEbiography}
\vspace{-1cm}
\begin{IEEEbiography}[{\includegraphics[width=1in,height=1.25in,clip,keepaspectratio]{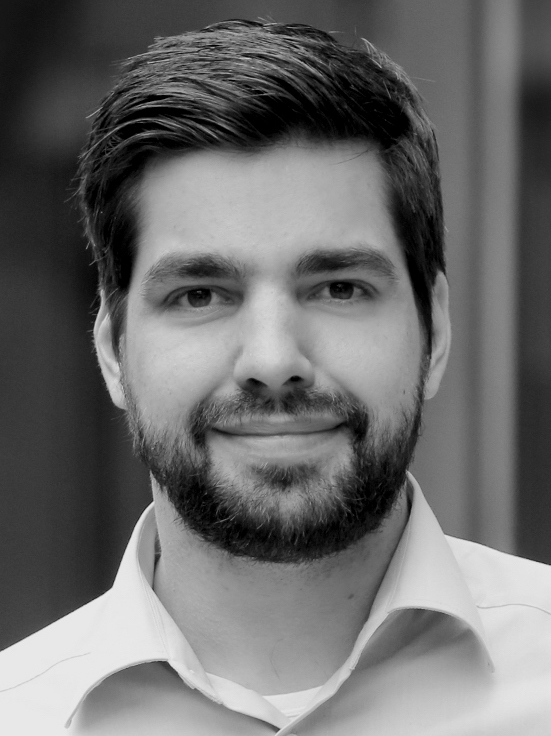}}]{Bastian Alt}
received his B.Sc. and M.Sc. degrees in electrical engineering and information technology from Technische Universität Darmstadt in November 2013 and December 2016, respectively. Since January 2017, he is working as a research associate within the Self-Organizing Systems Lab at Technische Universität Darmstadt.  In January 2022, he received the doctoral degree (Dr.-Ing.) at the department of electrical engineering and information technology, Technische Universität Darmstadt. He is interested in the modelling of network systems using tools from machine learning, control and optimization.
\end{IEEEbiography}
\vspace{-1cm}
\begin{IEEEbiography}[{\includegraphics[width=1in,height=1.25in,clip,keepaspectratio]{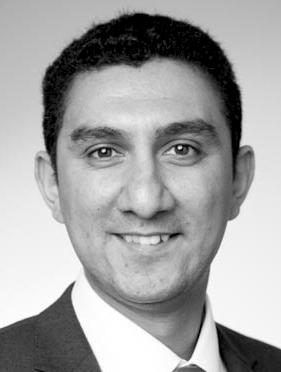}}]{Amr Rizk}
received the doctoral degree (Dr.-Ing.) from the Leibniz Universität Hannover, Germany, in 2013. After that he held postdoctoral positions at University of Warwick, UMass Amherst and the TU Darmstadt, Germany. From 2019 to 2021 he was an assistant professor at Ulm University, Germany. Since 2021 he is a professor at the department for computer science at the University of Duisburg-Essen, Germany. He is interested in performance evaluation of communication networks, stochastic models of networked systems and their applications.
\end{IEEEbiography}
\vspace{-1.0cm}
\begin{IEEEbiography}[{\includegraphics[width=1in,height=1.25in,clip,keepaspectratio]{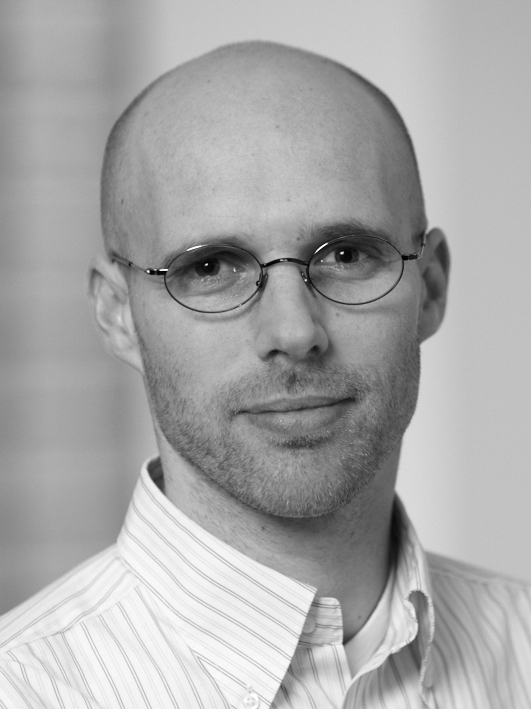}}]{Heinz Koeppl}
received the M.Sc. degree in physics from Karl-Franzens University Graz, in 2001, and the PhD degree in electrical engineering from the Graz University of Technology, Austria, in 2004. After that he held postdoctoral positions at UC Berkeley and Ecole Polytechnique Federalede Lausanne (EPFL). From 2010 to 2014 he was an assistant professor with the ETH Zurich and part-time group leader at IBM Research Zurich, Switzerland. Since 2014 he is a full professor with the Department of Electrical Engineering and Information Technology, Technische Universität Darmstadt, Germany. 
He received two awards for his PhD thesis, the Erwin Schrödinger Fellow-ship, the SNSF Professorship Award and currently holds an ERC consolidator grant. He is interested in stochastic models and their inference in applications ranging from communication networks to cell biology.
\end{IEEEbiography}

\end{document}